\documentclass[a4paper,10pt]{scrartcl}%Autor: Bernhardt, Ge Qu

\usepackage{actuarialsymbol}%actuarial notation
\usepackage{amsmath}%many things (e.g. equation in theorem command, align, gather ...)
\usepackage{amssymb}%R, C, Q with double-bar, {\setminus} otherwise strange spacing depending on sets like $R_+\setminus D$ and $[0,T[\,\setminus D$
\usepackage{amsthm}%\newtheorem command
\usepackage{array}% for \newcolumntype and > operator and other table tools
\usepackage{bbm}%indicator function (1 with double-bar, missing in amssymb)
\usepackage[
    backend=biber, % using helper program for biblatex (often not needed)
    articlein=false, % removes 'In:' from articles, only available for ext-styles
    dashed=false, % removes '-' and displays author names when two articles are written by same authors
    giveninits=true,% abbreviated first names
    maxcitenames=2, % et al when there are more than two names in citing
    natbib=true, % macros from incompatible package natbib like 'citep' or 'citet'
    style=ext-authoryear-ibid, % yields author (year) in Reference list
    uniquename=init,% technical commands to avoid warning (ext-authoryear-ibid uses full name while giveninits reduces full name)
    ]{biblatex}%bibliography creator and reference list changes
\usepackage{bigstrut}%for macro \bigstrut to create more cell space in tabular environments    
\usepackage{caption}%creating captions for any environment
\usepackage{chngcntr}%for figures, counter within section
\usepackage{enumitem}%tools to change list environments
\usepackage[margin=2.5cm]{geometry}%changing page layout, 3cm edge, symmetric
\usepackage{graphicx}%to include graphics
\usepackage[mathlines]{lineno}% creates numbers for lines, useful for peer review
\usepackage{setspace}% Chicago Style as required by NAAJ
\usepackage[table]{xcolor}%alternating colors for tables

\allowdisplaybreaks%breaks up math environments over pages otherwise only on the same page
\citetrackerfalse 
\newtheoremstyle{NoItalic}{}{}{}{}{\bfseries}{.}{.5em}{\thmname{#1}\thmnumber{ #2}\thmnote{ #3}}\theoremstyle{NoItalic}

\newtheorem{defi}{Definition}[subsection]
\newtheorem{ex}[defi]{Example}
\newtheorem{lem}[defi]{Lemma}
\newtheorem{pro}[defi]{Proposition}
\newtheorem{rmk}[defi]{Remark}
\newtheorem{simu}[defi]{Simulation}
\newtheorem{thm}[defi]{Theorem}
\makeatletter
\let\c@lem=\c@figure
\makeatother

\makeatletter 
\renewenvironment{proof}[1][\proofname]{\par
\pushQED{\qed}%
\normalfont \topsep6\p@\@plus6\p@\relax
\trivlist
\item\relax
{\itshape
#1\,\normalfont\@addpunct{:}}\hspace\labelsep\ignorespaces
}{%
\popQED\endtrivlist\@endpefalse
}
\makeatother

\DefineBibliographyStrings{english}{and={\&}} % replacing 'and' with '&'
\DeclareFieldFormat[article,incollection]{title}{#1} % removes quotation marks for title in article
\DeclareFieldFormat[article,periodical]{number}{\mkbibparens{#1}} % volumn numbers are in round brackets 
\DeclareNameAlias{sortname}{family-given} % surname-firstname in bibliography
\begin{document} 

\title{Wealth heterogeneity in a closed pooled annuity fund}
\author{Thomas Bernhardt\footnote{University of Manchester, {\texttt{thomas.bernhardt@manchester.ac.uk}}} \hspace{0em} and Ge Qu\footnote{University of Michigan, {\texttt{quge@umich.edu}}}}
%\date{}
\maketitle

\vspace{-4em}

\section*{\begin{center}Abstract\end{center}}
\begin{abstract}
\noindent The stability of income payments in a pooled annuity fund is studied. In those funds, members receive a fluctuating income depending on their experienced mortality in exchange for their pension savings. The focus is on describing the influence of different initial savings on the ability of the fund to provide a stable income in retirement. Because of this, members coincide in their characteristics except for their initial savings. We identify a term, which we dub ``implied number of homogeneous members'', that directly links the initial savings to the size of the income fluctuations. Our main contribution is the analysis of this term and the development of a criterion to answer the question of whether or not a given group of same-aged people should pool their funds together.
\\[1em]
\textbf{Keywords:} Tontine; unsystematic risk; wealth heterogeneity; the implied number of homogeneous members; beneficial groups.
\\[0.5em]
\textbf{JEL codes:} C61, G22
\end{abstract}

%---------------------------
\section{Introduction}\label{section:Intro}
%---------------------------

Even though issues with defined contribution schemes have been known for years, see \textcite{OECD2012}, most \textcite{OECD2019} countries continue to shift from pensions to defined contribution schemes. The increasing costs of guarantees on income in retirement have led to that shift. The rising costs are due to all-time high life expectancy and unforeseen long periods of low-interest rates. 

Governments and societies around the world ask for new retirement products. In the UK, the opposition in the \textcite{HOC2018} has questioned the government on the lack of innovative products to convert a lump sum into a lifelong retirement income. In Canada, the \textcite{HOOPP2018} is concerned with a lack of products that could pool mortality risk and missing default options that cost-effectively decumulate retirement funds. \textcite{Treas2018} of Australia has created a list of checkpoints for new retirement products: 1st high expected income and low fees, 2nd income stability, 3rd access to underlying funds, and 4th death and reversionary benefits. 

In this context, pooled annuity funds, or tontines in the academic literature, have gained lots of attention. They remove costly guarantees by adjusting the retirees' income payments with experienced mortality rates. Hence, pooled annuity funds have a built-in way to achieve the 1st point of the list, low fees. However, the operation of pooled annuity funds is still a matter of research. This paper is a stepping stone in developing a mathematical theory for pooled annuity funds.

We focus on the 2nd point of the list, income stability. In pooled annuity funds, retirees join a pool of members to mitigate the uncertainty about their lifetime (called idiosyncratic risk). Members who live shorter than expected fund the income of members who live longer than expected. However, only over time this knowledge becomes evident and causes members' income payments to fluctuate. When members are interchangeable (the homogeneous case), the uncertainty is only in members' death times rather than who dies. But, when members are noninterchangeable (the heterogeneous case), the order in which they die also impacts the reallocation of funds. Most notable is the situation when members contribute different savings amounts to the pool. There might be members with life savings of \pounds 100k and others with a \pounds 1m. Thus, savings could differ by orders of magnitudes yielding vastly different amounts available for redistribution over time depending on the order of deaths.

In this paper, we identify a term that drives the instability in the income fluctuations coming from the initial savings distribution of its members, which we dub ``implied number of homogenous members''. An easy way to access it is by computing the variance of the random income payments, which has been done in the literature many times, e.g.\ \textcite[Eq.24]{DoGuNi2014} or \textcite[Eq.13]{FoSa2016} or \textcite{Wright2018}. However, to our knowledge, it has never been studied as a tool to understand income stability. We use the implied number to explain phenomena that can be found numerically and, most importantly, develop a criterion to answer the question of whether or not it is beneficial for a group to pool their funds together.

We numerically observe the following statements about a group which is heterogeneous in the initial savings but homogeneous in all other characteristics. We show that those statements hold for any such group by studying the implied number.
\begin{enumerate}[label=(\alph*)]
    \item[(i)] 
        Different initial savings always negatively affect the income stability of the fund,
    \item[(ii)] 
        wealthy members always increase the stability of their income payments when they pool their funds with poorer members,
    \item[(iii)]
        if the highest savings amount is at most 2 times the lowest amount, then all members benefit from pooling their funds together, and the income stability is close to the homogeneous case.
    \item[(iv)]
        if the highest savings amount is more than 2 times the lowest amount, then poor members tend to prefer no wealthy members unless there are comparable many wealthy members.       
\end{enumerate}
In particular, we demonstrate observation (iv) by constructing groups in which some members would have a higher implied number (equivalent to increased stability) when they leave the group and form their own smaller pool.

We call a pool beneficial if the whole group yields the maximal implied number under all subsets of members. We introduce this notion to avoid situations like in observation (iv). In a beneficial group, every member benefits in terms of stability from pooling their funds, and nobody wants to change. In particular, it is not in the best interest of all members to pool their funds together in a non-beneficial pool. Some members would prefer to leave the group and form their own pooled annuity fund to increase the stability of their income payments. We determine whether or not a pool is beneficial by identifying all its beneficial subgroups. We show that the cumulative union of members with increasingly higher savings is a sequence of subgroups that contains all beneficial ones. We imagine actuaries using this outcome of our research to decide on savings limits for participation in newly established pooled annuity funds.

Our research is quite different from the rest of the existing literature. Most of the literature addresses specifics of how to share contributed funds of deceased members among alive members in the pool. In particular, the concept of actuarial fairness, i.e.\ that each member receives their contribution on average back, has a lot of attention. However, this concept appears to be debatable. Most intuitive sharing rules fail to be actuarially fair, even though they are close to being so, see \textcite{Donnelly2015}. And, the known sharing rules that are actuarially fair have caveats. For example, some rules equate the value of money to alive members with money to dead people yielding income payments to deceased members instead of funding the income of surviving members, see \textcite{DoGuNi2014}. Other actuarially fair rules have properties that render them useless for pooled annuity funds, like the rule by \textcite{Denuit2019}, called the conditional mean risk-sharing rule. \textcite[Example 4.1]{DeHiRo2022} demonstrates that the conditional mean risk-sharing rule is hypersensitive towards small changes in the initial savings with possible no mitigation of lifetime uncertainty in extreme cases. The current literature sees actuarial fairness as a promise to ensure that we can pool people with arbitrary characteristics together. But in fact, we show in this paper that the actuarially fair rule by \textcite{DoGuNi2014} suffers from the same issues related to different initial savings as our pooled annuity fund. Beyond actuarial fairness, some researchers use utility approaches from optimal control to determine income payments to members and never consider funds separated between members, see \textcite{MiSa2015}. Unfortunately, this research yields cumbersome or only numerically accessible formulas so far. We redirect the interested reader to \textcite{ChHiKl2019} and \textcite{ChRaSe2020}. Similar inaccessible, \textcite{Sabin2010} considers a system of linear equations to ensure that no individual has an advantage over another in a pooled annuity fund. But he is unable to present a solution to the system. We avoid all of the above complications and choose the simple and arguably most intuitive rule by \textcite{PiVaDe2005}, which splits the funds of deceased members evenly among survivors according to their share in the fund.
 
Closest to our approach are the works of \textcite{BeDo2021} and \textcite{QiSh2013}. Both study the income stability of a pooled annuity fund over time. \textcite{QiSh2013} do a numerical study with real-life features like systematic risk, while \textcite{BeDo2021} approach the problem analytically. Their focus is on the minimal number of members in the pool to mitigate the lifetime uncertainty efficiently. Particular interest to us is how \textcite{BeDo2021} quantify lifetime uncertainty in their homogeneous pooled annuity fund. They introduce bounds and ask how long and with what probability members in the pool receive an income within the bounds. We use their approach. However, we are less interested in the number of members to achieve a targeted stability. Also, we are less interested in modelling the current world accurately. For example, the correlation between initial savings and mortality is of no concern to us. Instead, we are interested in the impact of different initial savings on the pooling mechanism. Our setting is rich enough to see its impact. Even though we make strong assumptions about the remaining lifetimes and initial savings (independent and identical distributed), our setting resembles the situation of a newly established pooled annuity fund with one cohort and one mortality distribution. Practitioners could use our results to decide on maximal savings limits for newly established pooled annuity funds.

The main focus of our paper, the ``implied number of homogenous members'', describes the impact of different initial savings on income stability regardless of the number of members in the pool.
Several authors identify the ``implied number of homogenous members'' as a tool in different contexts. \textcite{DoGuNi2014} use it to measure the relative attractiveness of two retirement products in terms of break-even costs. \textcite{FoSa2016} have found it when computing variances for payments for pooled annuities in a one-period setting. And the practitioner \textcite{Wright2018} suggests using a similar term to inform trustees about the risks involved in closed definite benefit schemes. But differently, we focus on the specific aspect of income stability affected by different initial savings and try to extract as much general information as possible. 

The structure of the rest of the paper is as follows. In Section 2, we introduce the pooled annuity fund, explain how we quantify the uncertainty about remaining lifetimes (idiosyncratic mortality risk), and demonstrate the issues that arise with different initial savings in a numerical study. In Section 3, we identify the ``implied number of homogeneous members'' as the main driver for the income instability caused by different initial savings from formulas that measure the idiosyncratic mortality risk. In Section 4, we study the ``implied number of homogeneous members'' and introduce the concept of a ``beneficial group'' as a criterion to determine whether or not a group should pool their funds together. Section 5 concludes the paper with a summary.
 
%-------------------
\section{Setting and problem}\label{section:setting}
%-------------------

%-------------------
\subsection{The pooled annuity fund}\label{section:operation}
%-------------------

We consider a group of $\,N\geq1\,$ individuals who form the entire membership of the pooled annuity fund at time 0. Each member is aged $x$ and is assumed to be an independent and identical copy of the rest. Each member $i$ has its own initial savings amount $s_i$. 

No other person joins the pool after time 0, and each member stays in the fund until death. The survival rates of the whole group are assumed to be known, which means there is no systematic mortality risk. Moreover, we take for granted that the survival rates are continuous and strictly decreasing until the limiting age, possibly $\infty$. There is a financial market in which the pooled annuity fund invests. We assume that the market yields a fixed interest rate $\,R\geq0$, which means no systematic investment risk.

%-------------------
\subsubsection{Future lifetime random variables and random initial savings}
%-------------------- 

We denote by $T_i$ and $s_i$ the remaining future lifetime and the initial savings of the $i$th member and assume that the member is alive and has savings initially, i.e.\ $T_i>0\,$ and $\,s_i>0$. Both are random variables that exist on an underlying probability space $(\Omega,\mathcal{F},\mathbb{P})$. 

At time 0, we know the savings $s_i$, while the remaining future lifetime $T_i$ is unknown. However, we insist that $s_i$ is random because we want to restrict the possible sequences $(s_i)_{i=1}^N$ for $\,N\uparrow\infty$ to use asymptotic results from probability theory. In the following, all probabilities and expectations are conditional given the information about the savings $(s_i)_{i=1}^N$, unless we specifically mention otherwise. Thus, we treat the savings $(s_i)_{i=1}^N$ as given constants.

We assume that $(T_i)_{i=1}^N$ and $(s_i)_{i=1}^N$ are independent sequences of independent and identical distributed random variables, along with $s_i$ having a finite fourth moment. We make those assumptions to apply the existing literature about sequences of random variables in favour of the least restrictive assumptions because this is not the focus of this paper.

%--------------------
\subsubsection{The income calculation}\label{subsubsection:income-calculation}
%-------------------

Every member $i$ of the pooled annuity has an individual fund account $W_i$. The value of the fund account changes over time due to investment returns, income withdrawals, and longevity credits, the latter coming from the reallocation of the fund accounts of deceased members. The fund account corresponds initially to the individual's savings, stays positive as long as the individual is alive, and will be re-allocated to other members after the individual passes away, i.e.\ $W_i(0)=s_i$, $\,W_i(t)>0\,$ for all $\,t<T_i$, and $\,W_i(t)=0\,$ for all $\,t\geq T_i$.

The pooled annuity fund is assumed to pay an income to alive members from their fund accounts periodically after a fixed unit of time, for example, a day, a week, a month, or a year at integer times, i.e.\ $\,t=0,1,2,3,\,$etc. The income is computed based on fair life annuity prices that use $\,R\geq0\,$ as the interest rate. Thus, the $i$th member receives at time $t$ the income payment
\begin{equation}\label{eq:C_i}
    C_i(t)=\frac{W_i(t)}{\ddot{a}(x+t)},
\end{equation}
in which 
\begin{align}
    \label{eq:ddota}\ddot{a}(x+t)&= 1+\sum_{\delta=1}^{\infty}(1+R)^{-\delta}\actsymb[\delta]{p}{x+t},
    \\\label{eq:p_x+t}\actsymb[\delta]{p}{x+t}&=\mathbb{P}\big[ T_i>t+\delta\,\big|\,T_i>t\big].
\end{align}

Recall that we assume that the survival rates $\actsymb[\delta]{p}{x+t}$ are continuous and strictly decreasing in $\delta$ until the limiting age, possibly $\infty$.

Observe that $\actsymb[\delta]{p}{x+t}$ is independent of $i$ due to the i.i.d.\ assumption on $(T_i)_{i=1}^N$. Also, note that the annuity prices depend on future investment returns and mortality rates, but they are deterministic in our setting because we exclude systematic risks.  

The pooled annuity fund pays an income only to alive members because we re-allocate the funds of deceased members, i.e.\ $W_i(t)=0\,$ for all $\,t\geq T_i$, implying $\,C_i(t)=0\,$ for all $\,t\geq T_i$.

%---------------------
\subsubsection{The longevity credit calculation}
%---------------------

Every member $i$ invests in the financial market with a fixed interest rate $\,R\geq0$. The value of the fund account grows from $\,W_i(t)-C_i(t)\,$ to $\,(W_i(t)-C_i(t))(1+R)\,$ between income payments at times $t$ and $t+1$. By definition, $W_i$ drops to zero when the $i$th member passes away. However, the pooled annuity fund retains the investment of $\,W_i(t)-C_i(t)\,$ from time $t$ and receives $\,(W_i(t)-C_i(t))(1+R)\,$ at time $t+1$, assuming $\,T_i\in(t,t+1]$. In particular, the funds of recently deceased members who have passed away within $(t,t+1]$ is
\begin{equation}\label{eq:D}
    D(t+1)=\sum_{k=1}^N(W_k(t)-C_k(t))(1+R)\mathbbm{1}_{\{T_k\in(t,t+1]\}},
\end{equation}     
in which $\mathbbm{1}_A$ is the indicator function of the set $\,A\subset\Omega$. 

The pooled annuity fund is assumed to increase the fund accounts of alive members with money from recently deceased members periodically after a fixed unit of time before members receive their income at integer times $\,t=1,2,3,\,$etc. The increases are called longevity credits. Alive members receive funds from deceased members based on their relative ownership of funds of living members. Thus, the fund of a member $i$ aged $\,x+t+1\,$ increases by
\begin{equation}\label{eq:M_ioriginal}
    M_i(t+1)=\frac{(W_i(t)-C_i(t))(1+R)\mathbbm{1}_{\{T_i>t+1\}}}{\sum_{k=1}^N(W_k(t)-C_k(t))(1+R)\mathbbm{1}_{\{T_k>t+1\}}}D(t+1).
\end{equation} 
Note that the denominator in (\ref{eq:M_ioriginal}) is strictly positive as long as members are alive at $t+1$. To see this, observe that (\ref{eq:C_i}) implies that $\,W_k(t)-C_k(t)=W_k(t)(1-1/\ddot{a}(x+t))\geq0$, which is not zero when $\,T_k>t+1\,$ because $\,W_k(t)>0\,$ and $\,\ddot{a}(x+t)>1\,$ in this case. If all members pass away, the pooled annuity fund stops paying longevity credits, and we can use the convention $\,0/0=0$. In particular, we can use the arithmetic operations of the extended real number line with $\,1/0=\infty\,$ and $\,0\cdot\infty=0\,$ to use (\ref{eq:M_ioriginal}) for all $t$, no matter if the $i$th member is alive or dead. 

If all members pass away, (\ref{eq:M_ioriginal}) is zero. But, (\ref{eq:D}) might be strictly positive. In this case, we choose to give the funds of the last deceased members to their heirs, i.e.\ we allocate to someone the remaining funds of the pool after all members have passed away. We introduce this rule for completeness only. The situation of no one left in the pooled annuity fund is irrelevant to our analysis.

The pooled annuity fund re-allocates funds only to alive members due to the indicator function of the set $\{T_i>t+1\}$ in the nominator of $M_i$, i.e.\ $M_i(t+1)=0\,$ for all $\,t+1\geq T_i$. Moreover, the pooled annuity fund re-allocates only exactly the funds of recently deceased members who have passed away in $(t,t+1]$ because of $\,\sum_{i=1}^NM_i(t+1)=D(t+1)\,$ if the denominator in (\ref{eq:M_ioriginal}) is positive.

%---------------------
\subsubsection{Recursive computation}
%---------------------

Each member $i$ withdraws an income from their fund account at the beginning of the time point $t$ and invests the remaining funds into the financial market until the end of the period at time $t+1$. Each alive member receives potentially longevity credits from recently deceased members at time $t+1$. In particular, the value of the fund account $W_i$ before the next income payment fulfils the recursive relationship 
\begin{equation}\label{eq:W_i}
    W_i(t+1)=[(W_i(t)-C_i(t))(1+R)+M_i(t+1)]\mathbbm{1}_{\{T_i>t+1\}}
\end{equation} 
with $\,W_i(0)=s_i$.

Note that $\,W_i(t+1)>0\,$ when $\,W_i(t)>0\,$ and $\,T_i>t+1$. To see this, observe that (\ref{eq:C_i}) implies $\,W_k(t)-C_k(t)=W_k(t)(1-1/\ddot{a}(x+t))>0\,$ and $\,M_i(t)\geq0$. In particular, the recursion (\ref{eq:W_i}) ensures $\,W_i(t)>0\,$ for all $\,t<T_i$ and $\,W_i(t)=0\,$ for all $\,t\geq T_i$, as stated at the beginning of Subsection \ref{subsubsection:income-calculation}. 

%---------------------
\subsubsection{Explicit formulas}
%---------------------

The particular choices for $C_i$ and $M_i$ in (\ref{eq:C_i}) and (\ref{eq:M_ioriginal}) preserve the wealth heterogeneity of the group over time in the way that the ratio of the value of the fund accounts of two alive members stays constant, i.e.\ $W_i(t)/W_j(t)=s_i/s_j\,$ for all $\,t<\min\{T_i,T_j\}$. See Lemma \ref{lem:W_i/W_j=s_i/s_j} for a proof in the Appendix. That means any fluctuation in the fund account of one member occurs in the fund account of all alive members. For example, a $3\%$ change for one member means a $3\%$ change for all members. It follows that there is a stochastic process that describes the fluctuations in the fund accounts for all members up until they die. The same is true for the income payments due to the linear relationship between income and wealth. More precisely, at integer times $\,t=0,1,2,\,$etc:
\begin{align}
    \label{eq:W_i(t)=s_i}W_i(t)&=\mathbbm{1}_{\{T_i>t\}}s_i\frac{\ddot{a}(x+t)}{\ddot{a}(x)}\frac{\actsymb[t]{p}{x}}{\actsymb[t]{\hat{p}}{x}},
    \\\label{eq:C_i(t)=C_i(0)}C_i(t)&=\mathbbm{1}_{\{T_i>t\}}C_i(0)\frac{\actsymb[t]{p}{x}}{\actsymb[t]{\hat{p}}{x}},
\end{align}
in which
\begin{equation}\label{eq:tphatx}
    \actsymb[t]{\hat{p}}{x}=\frac{\sum_{i=1}^Ns_i\mathbbm{1}_{\{T_i>t\}}}{\sum_{i=1}^Ns_i}.
\end{equation}
See Proposition \ref{pro:C_i(t)=C_i(0)} for a proof in the Appendix.

%--------------------
\subsection{Quantifying idiosyncratic mortality risk}\label{subsection:quantifying}
%--------------------
We assume that $\actsymb[\delta]{p}{x+t}$ in (\ref{eq:p_x+t}) are the ideal survival rates. We think of $T_i$ as drawn from a distribution with those rates. However, the finite sample $(T_i)_{i=1}^N$ may fail to reproduce the survival rates. For example, if we keep drawing samples of $(T_i)_{i=1}^N$, then sometimes all $T_i$ from $i$ equals 1 to $N$ will be below the median of the distribution. In particular, we would be unable to identify the median from those instances. That is called sample error in statistics and idiosyncratic mortality risk in life insurance. It should not be confused with systematic mortality risk, i.e.\ the issue of predicting the ideal survival rates.

Members of the pooled annuity fund receive a random income. The randomness stems from the idiosyncratic mortality risk as we excluded systematic risks. In our specific model, the difference between the ideal survival rates $\actsymb[t]{p}{x}$ and the experienced weighted survival rates $\actsymb[t]{\hat{p}}{x}$ causes fluctuations in the income payments, see equation (\ref{eq:C_i(t)=C_i(0)}). 

The fluctuations in the income payments disappear with an increasing number of members in the pooled annuity fund. Even more, every member would receive a constant income for life if there were infinitely many members in the pooled annuity fund. Indeed, we have chosen the income payments $C_i$ in (\ref{eq:C_i}) in connection with the longevity credit $M_i$ in (\ref{eq:M_ioriginal}) such that this holds. It is a consequence of the law of large numbers applied to the independent i.i.d.\ sequences $(T_i)_{i=1}^N$ and $(s_i)_{i=1}^N$. More precisely, without conditioning on $(s_i)_{i=1}^N$,
\begin{equation*}
    \actsymb[t]{\hat{p}}{x}=\frac{N}{\sum_{i=1}^Ns_i}\frac{\sum_{i=1}^Ns_i\mathbbm{1}_{\{T_i>t\}}}{N}\xrightarrow[]{N\uparrow\infty}\frac{\mathbb{E}[s_1\mathbbm{1}_{\{T_i>t\}}]}{\mathbb{E}[s_1]}=\frac{\mathbb{E}[s_1]}{\mathbb{E}[s_1]}\mathbb{E}[\mathbbm{1}_{\{T_i>t\}}]=\actsymb[t]{p}{x}.
\end{equation*}
In particular, we can quantify the idiosyncratic mortality risk in the pooled annuity fund by studying the stability of the income payments.

It is always possible that no member dies for some time or that all members die suddenly. Thus, there will always be some uncertainty about the income stability for any fixed time interval. More technically, we assume that the survival probability $\actsymb[t]{p}{x}$ is continuous in time $t$, which is the case for most mortality models like the Gompertz-Makeham law of mortality. But, $\,t\mapsto\actsymb[t]{\hat{p}}{x}$ is a discontinuous step function, and therefore, the ratio $\,\actsymb[t]{p}{x}/\actsymb[t]{\hat{p}}{x}\,$ is hardly $1$. As we cannot avoid fluctuations, we tolerate fluctuations within some bounds. We use the following definition, which originates from \textcite{BeDo2021}. 

\begin{defi}\label{defi:P>beta}
    $ $
    \begin{enumerate}[label=(\alph*)]
        \item 
            We call the income of the pooled annuity fund
            \begin{itemize}
                \item[(i)] 
                    stable within the bounds $\,\varepsilon_2>0\,$ and $\,\varepsilon_1\in(0,1)\,$ until time $\,t=0,1,2,\,$etc.\ if the income payments for any member $i$ lie within $\,[C_i(0)(1-\varepsilon_1),C_i(0)(1+\varepsilon_2)]\,$ until $\,\min\{T_i,t\}$,
                \item[(ii)]
                    strongly stable within the bounds $\,\varepsilon_2>0\,$ and $\,\varepsilon_1\in(0,1)\,$ until time $\,t\in[0,\infty)\,$ if the ratio $\,\actsymb[\delta]{p}{x}/\actsymb[\delta]{\hat{p}}{x}\,$ stays within the interval $\,[1-\varepsilon_1,1+\varepsilon_2]\,$ for all $\,\delta\leq t$.
            \end{itemize}
        \item
            We say that the income is (strongly) stable within the bounds $ \,\varepsilon_1,\varepsilon_2$ until time $t$ with certainty $\,\beta\in[0,1]\,$ if it is strongly stable until time $t$ in at least $100\beta\%$ of all future scenarios, i.e.\
            \begin{equation}\label{eq:P>beta}
                \mathbb{P}\Big[1+\varepsilon_2\geq\frac{\actsymb[\delta]{p}{x}}{\actsymb[\delta]{\hat{p}}{x}}\geq 1-\varepsilon_1\quad\mbox{for all $\,\delta\leq t$}\Big]\geq\beta
            \end{equation}
            holds.
    \end{enumerate}
\end{defi}

If any of (a) holds, every member who passes away before time $t$ receives an income for life that increases by at most $100\varepsilon_2\%$ and decreases by at most $100\varepsilon_1\%$ compared to their initial income level. Any member who survives until time $t$ receives a fluctuating income with the same properties until time $t$ and owns funds at time $t$ that are worth an annuity priced according to $(\ref{eq:ddota})$ of their current income.

The difference between (i) and (ii) is when we want the income to be within the bounds $\varepsilon_1,\varepsilon_2$. Condition (i) only requires the income to be within the bounds when members receive income payments. Condition (ii) requires the income process in equation (\ref{eq:C_i(t)=C_i(0)}) to be within the bounds for all times. The advantage of (ii) is that we can study our research questions independent of any mortality distribution in the following Sections. We could interpret (ii) as a prudent insurance company that observes the income stability of the fund and would act immediately if the income reached the bounds and would not wait to pay an amount outside the bounds.

Our analysis focuses on the prior duration that a pooled annuity fund can support a stable income for its members, i.e.\ Definition \ref{defi:P>beta}$(b)$. Stability cannot be guaranteed in all scenarios, even when we tolerate fluctuations within bounds. The possibility that no member dies or all members die suddenly shows that the certainty of $\,\beta=1\,$ for given $\,t>0\,$ is unachievable for any bounds $\,\varepsilon_1,\varepsilon_2\,$ with trivial exceptions. However, $\,\beta<1\,$ yields a maximal time $\,t>0\,$ fulfilling \ref{defi:P>beta}$(b)$. The maximal time is relevant for the provider as well as the members of the pooled annuity fund. Both parties gain financial security about the future in an environment without guarantees. 

It is worth noting that pooled annuity funds do not last forever, and the maximal time determines when we expect other retirement products to take over. 

Figure \ref{fig:sample} illustrates Definition \ref{defi:P>beta}$(b)$. Notice that the change from stable to unstable happens before the process crosses the bounds $\,\varepsilon_{1,2}=10\%$. The shown sample of $\,\delta\mapsto\actsymb[\delta]{p}{x}/\actsymb[\delta]{\hat{p}}{x}\,$ is one of the 90\% future scenarios for which the income payments of all members are within the thresholds for at least $\,t=15.06\,$ years (the maximal time that fulfils (\ref{eq:P>beta}) based on the used life table). See Simulation \ref{simu:2groups} in the appendix for an account of how we calculate $\,t=15.06\,$ using a Monte Carlo simulation with $\,R=10^6\,$ samples. In particular, we expect an error of magnitude of $\,1/\sqrt{10^6}=0.1\%\,$ by the central limit theorem. 
\begin{center}
    \includegraphics[trim=0 10 0 30,width=0.45\linewidth]{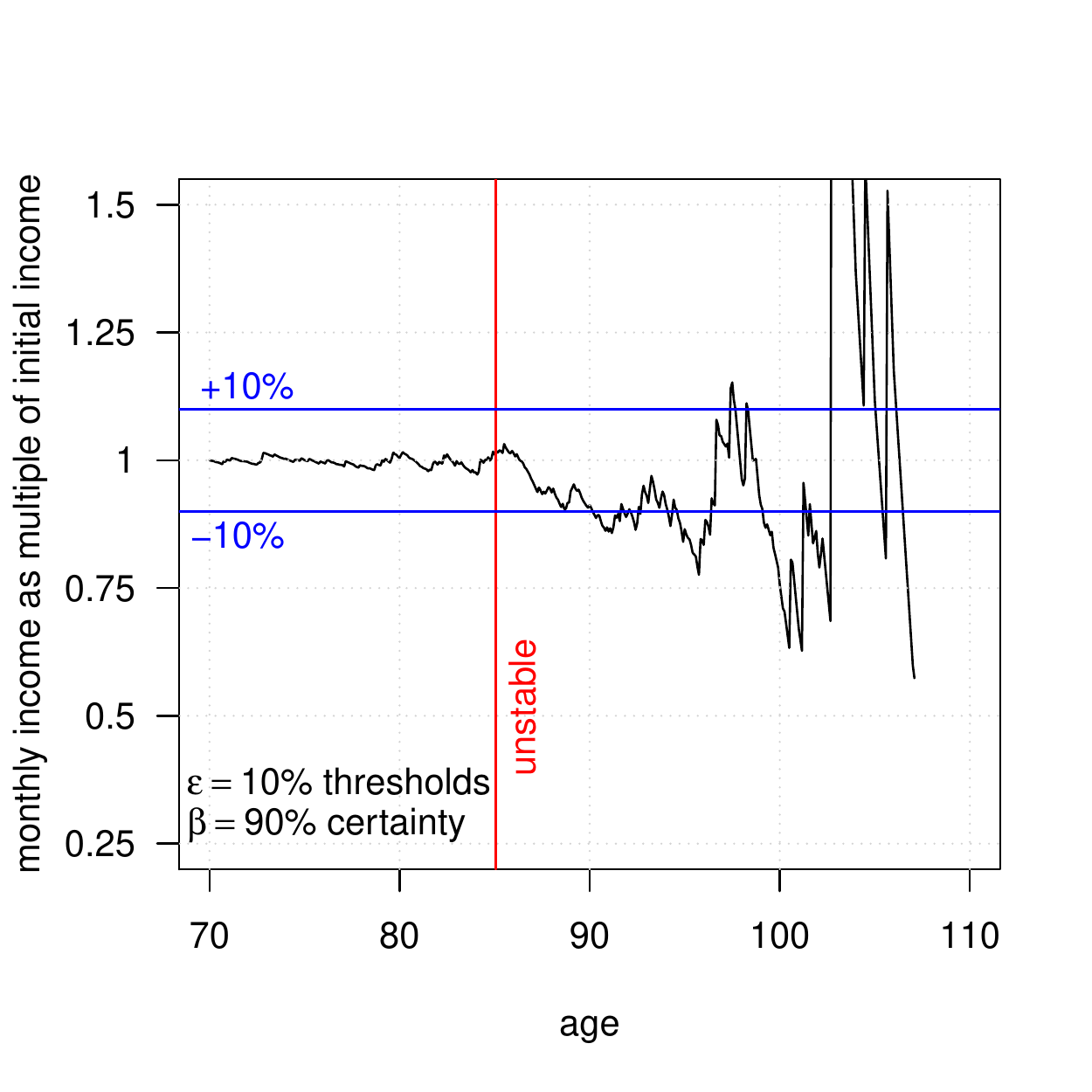}
    \captionof{figure}{A sample of the income fluctuations for all members in a pooled annuity fund in black. The pool initially consists of 1000 members, each 70 years old, of whom 900 have savings of \pounds1 and 100 have savings of \pounds10. The mortality distribution follows UK period data for both sexes from the \textcite{HMD2020UK}. The income lies within the 10\% thresholds in blue until the red line and is part of the 90\% possible outcomes with the same property.} 
    \label{fig:sample}
\end{center}

%--------------------
\subsection{The issue with wealth heterogeneity}\label{subsection:problem}
%--------------------

We study numerically the situation of a pooled annuity fund with 1000 members split into two subgroups. The first subgroup consists of $N_1$ members, while the second subgroup consists of $1000-N_1$ members. Members of each subgroup have the same initial savings amount, $m$ or $M$, respectively. Further, we assume that members from the second subgroup have more savings than members from the first subgroup, i.e.\ $m<M$, and refer to them as rich or poor. The exact values of the initial savings are irrelevant for the computations; only its ratio $\,m/M\,$ enters the calculations. We refer to $\,m/M\,$ as the income or wealth ratio. 

We are interested in the income stability of the fund when we vary the initial number of poor members $N_1$ and the income ratio $\,m/M$. Note that the total number of members in the pool is always 1000. Also, we are interested in how the mixed fund compares to the situation when we would keep the two subgroups in separate pooled annuity funds. Here, the number of members varies from 0 to 1000.

We use Monte Carlo simulation to find the maximal time $t$ that fulfils (\ref{eq:P>beta}) for given $N_1$ and $\,m/M$. We use $\,\varepsilon_1=10\%$ and $\,\beta=90\%\,$ for the stability parameters. Also, we choose $\,\varepsilon_2=\infty$, which removes the upper bound and simplifies our analysis in Section \ref{section:derivation}. See Simulation \ref{simu:2groups} in the appendix for an account of how we calculate the maximal time $t$ using a Monte Carlo simulation with $\,R=10^6\,$ samples. In particular, we expect an error of magnitude of $\,1/\sqrt{10^6}=0.1\%\,$ by the central limit theorem. 

First, we look at a slice of our numerical results to help the reader better understand the later graphs. Let us fix the number of poor members at $\,N_1=800$. Then, the number of rich members is $\,1000-N_1=200\,$ because the total number is 1000. We consider three possible pooled annuity funds: 1st the fund that consists of poor members, 2nd the fund that consists of rich members, and 3rd the fund that consists of all members. We are interested in the maximal time that each of the three pools provides a stable income to its members (to measure how well the pools diversify idiosyncratic risk). Table \ref{tab:Poor=800} shows our numerical findings depending on the income ratio $m/M$.
\begin{center}
\begin{tabular}{c|c|cccccc}
    group & size & $\tfrac{m}{M}=1$ & $\tfrac{m}{M}=0.7$ & $\tfrac{m}{M}=0.5$ & $\tfrac{m}{M}=0.3$ & $\tfrac{m}{M}=0.2$ & $\tfrac{m}{M}=0.1$\bigstrut
    \\\hline only poor & $800$ & 21.70 & 21.70 & 21.70 & 21.70 & 21.70 & 21.70
    \\only rich & $200$ & 15.41 & 15.41 & 15.41 & 15.41 & 15.41 & 15.41
    \\mixed & $1000$ & 22.57 & 22.48 & 22.15 & 21.24 & 20.12 & 18.48
\end{tabular}
    \captionof{table}{Maximal time in years for which a pooled annuity fund can provide a stable income with $\,\varepsilon_1=10\%\,$ and $\,\beta=90\%\,$ and varying income ratio $\,m/M$. Members are initially 70 years old and follow UK period data for both sexes from the \textcite{HMD2020UK}.}
    \label{tab:Poor=800} 
\end{center}
The income ratio $\,m/M$ does not influence the stability of the income payments when we separate poor from rich. Both pools consist entirely of homogeneous members. The pooled fund with more members has a higher maximal time of 21.70 years than the pool with fewer members and a maximal time of 15.41 years. Indeed, the law of large numbers predicts that increasing the number of members would decrease fluctuations and increase income stability. In contrast, the income ratio $\,m/M$ influences the income stability in the combined pool of all members. In our example, the maximal time monotonically changes from 18.48 years to 22.57 years when rich members have 10 times the savings of the poor members compared to the homogeneous case with 1000 members and no difference between rich and poor. But probably the most insightful information hides within the numbers between the different pools. The maximal time is longer for the fund consisting exclusively of all poor members at 21.70 years than for the mixed fund between 18.48 and 21.24 years (when rich members have at least 3 to 4 times more than poor members). So, the pool with fewer members diversifies idiosyncratic risk better than the pool with more members for low ratios $\,m/M$, which disputes the essence of the law of large numbers.

Next, we fix the income ratio at $\,m/M=0.3\,$ and vary the number of poor members $N_1$. Again, we consider three funds: 1st the fund that consists of poor members, 2nd the fund that consists of rich members, and 3rd the fund that consists of all members. Note that the number of poor members implies the number of members in each of the three pools because we fix the total number at 1000. More precisely, the 1st fund consists of $N_1$ members, the 2nd fund consists of $\,1000-N_1$ members, and the 3rd fund consists of 1000 members. Figure \ref{fig:3pools} shows our numerical findings depending on the number of poor members $N_1$.
\begin{center}
    \includegraphics[trim=0 10 0 30,width=0.45\linewidth]{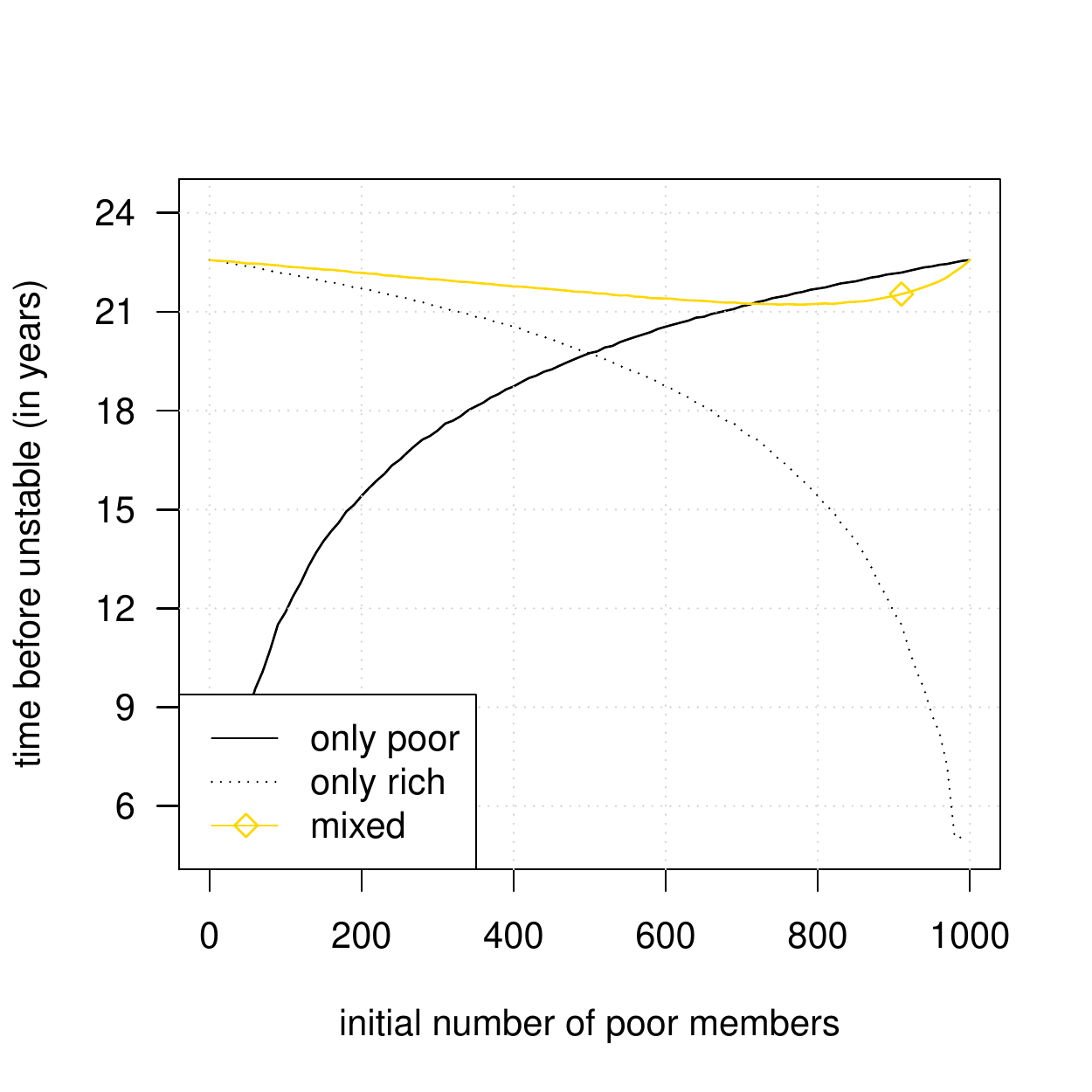} 
    \captionof{figure}{Comparing 1) the poor member pool, 2) the rich member pool, with 3) the mixed pool. The graphs show the maximal time in years of income stability with $\,\varepsilon_1=10\%$, $\,\beta=90\%$, $\,m/M=0.3$, and varying numbers of poor members $N_1$. Members are initially 70 years old and follow UK period data for both sexes from the \textcite{HMD2020UK}.}
    \label{fig:3pools} 
\end{center}
The number of poor members $N_1$ strongly influences the stability of the income payments when we separate poor members from rich. $N_1$ determines the number of members in both pools. Increasing the number of members increases the income stability as predicted by the law of large numbers. Both funds behave symmetrically to each other because both funds consist of homogeneous members. In contrast, the number of poor members $N_1$ moderately influences the income stability in the mixed pool. Here, the number of members is always 1000. The wealth distribution alone drives the changes in income stability in a non-monotonic way. The maximal time the mixed pool provides a stable income is always below the time of a homogeneous group with 1000 members. It is always above the time of a fund consisting exclusively of rich members. But, it depends on the precise number $N_1$ of poor members whether or not it is above or below the time of a pool consisting exclusively of poor members. If the proportion of poor members is large enough, then a smaller fund consisting of poor members only would yield a better diversification of idiosyncratic risk than the mixed fund.

Last, we present our full numerical results in Figure \ref{fig:Poor-Rich}, where we vary the number of poor members $N_1$ and the income ratio $\,m/M$.
\begin{center}
    \includegraphics[trim=0 10 0 30,width=0.45\linewidth]{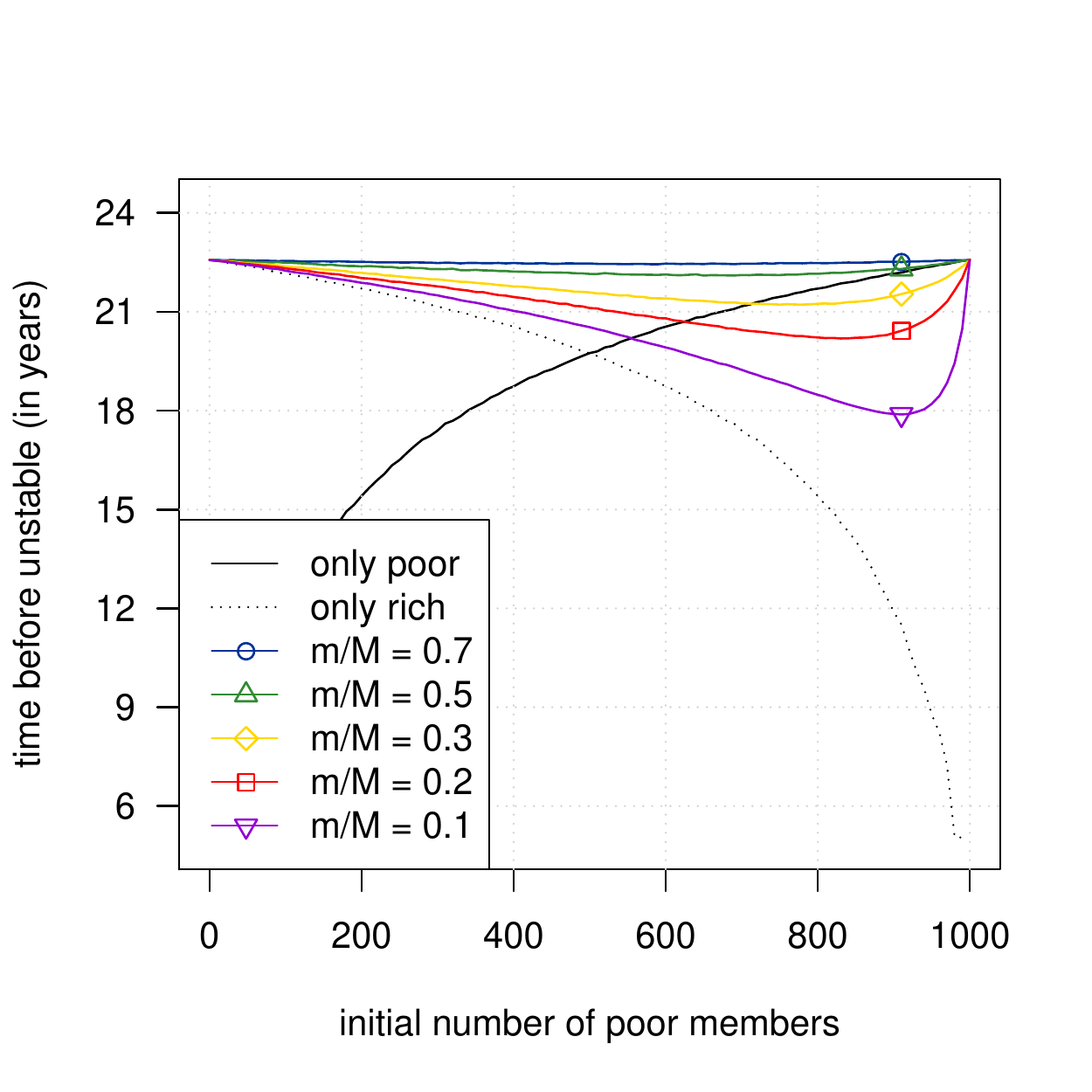} 
    \captionof{figure}{Maximal time for which a pooled annuity fund can provide a stable income with $\,\varepsilon_1=10\%\,$ and $\,\beta=90\%$. Pooling two subgroups with a fixed total member size of 1000 (coloured graphs) and separated subgroups with varying member sizes (dotted and black lines). The graphs show the impact of changing the initial number of poor members and the income ratio $\,m/M$ (coloured graphs only). Members are initially 70 years old and follow UK period data for both sexes from the \textcite{HMD2020UK}.}
    \label{fig:Poor-Rich} 
\end{center}
Table \ref{tab:Poor=800} is a slice of the results of Figure \ref{fig:Poor-Rich} when the x-value equals 800, while Figure \ref{fig:3pools} contains three of the seven graphs of Figure \ref{fig:Poor-Rich}. New to us is the slightly hidden qualitative change of the graphs for income ratios above and below $\,m/M=0.5$. The time a mixed fund provides a stable income for income ratios above 0.5 is above the time of a fund that exclusively consists of poor members (different from our previous observation in Figure \ref{fig:3pools}). In this case, the maximal time is close to the maximal time of a group with 1000 homogeneous members. In contrast, whether or not the maximal time is above or below the maximal time of the group of poor members depends on the number $N_1$ of poor members when the income ratio is below 0.5.

We summarise our findings in the following remark.
\begin{rmk}\label{rmk:graph}
Detailed analysis of Figure \ref{fig:Poor-Rich}:
    \begin{enumerate}[label=(\alph*)]
    \item 
        All graphs are below 22.57 years, which is the maximal time of a pool with 1000 members having equal initial savings and following the used mortality table. Hence, wealth heterogeneity negatively affects the stability of the fund. 
    \item 
        All coloured curves are above the dotted black line, meaning members that are rich benefit consistently from increased stability by pooling risk with poor members.
    \item
        The black line crosses some coloured curves, i.e.\ poor members might be worse off in a larger heterogeneous pool than in a smaller homogeneous one. 
    \item 
        The coloured curves corresponding to income ratios above 0.5 are above the black line and close to 22.57 years, i.e.\ wealth heterogeneity has little effect when the richest member has at most double the funds of the poorest member. 
    \item    
        The coloured curves corresponding to income ratios below 0.5 are below the black line for large proportions of poor members in the pool. In contrast, they are above the black line for enough participation of rich members. Thus, if the percentage of poor members in the fund is large enough, then a smaller fund consisting of only poor members yields a better diversification than a mixed fund.
    \end{enumerate}
\end{rmk}

Most notably, whether or not a group of members should be together in terms of income stability in a pooled annuity fund depends on the initial distribution of savings. Heuristically, we can explain why increasing the number of members may decrease income stability. Higher savings imply more funds from deceased members and more funds released upon death. Thus, an additional individual who owns disproportionately more funds than the rest of the pool increases the instability in the income rather than decreasing it.

The main issue with wealth heterogeneity is whether or not a given group of people benefits in terms of income stability from being together in a pooled annuity fund. Moreover, we are interested in whether or not Remark \ref{rmk:graph}(a)--(e) holds in general. 

%------------------------
\section{Mathematical description}\label{section:derivation}
%------------------------

%------------------------
\subsection{Asymptotic statistic}\label{subsection:asymptotic}
%------------------------

It is possible to find a formula that approximates the graphs in Figure \ref{fig:Poor-Rich}. The i.i.d.\ assumptions about the remaining lifetimes and initial savings allow us to link a Gaussian process to the income fluctuations (Donsker's theorem). 

We start with some notation and the notion of a transformed time, which allows us to state the results independently from the underlying mortality distribution. Let $F$ be the distribution function of the remaining future lifetimes of the members in the pool and $\hat{F}$ its empirical version weighted by savings, i.e., for all $\,t\geq0$,
\begin{gather*}
    F(t)=1-\actsymb[t]{p}{x},
    \\\hat{F}(t)=1-\actsymb[t]{\hat{p}}{x}=\frac{\sum_{i=1}^Ns_i\mathbbm{1}_{\{T_i\leq t\}}}{\sum_{i=1}^Ns_i}.
\end{gather*}
We consider the sequence of transformed future lifetimes $\,(U_i)_{i=1}^N=(F(T_i))_{i=1}^N$. It is a sequence of i.i.d.\ uniform random variables. Here, we have used that $\,t\mapsto\actsymb[t]{\hat{p}}{x}\,$ is continuous. Moreover, $F^{-1}$ is a well-defined function because $\,t\mapsto\actsymb[t]{\hat{p}}{x}\,$ is strictly decreasing until the limiting age. Hence, the following terms have the same distribution regardless of the underlying mortality distribution:
\begin{align}
    \label{eq:F^-1px}\actsymb[F^{-1}(v)]{p}{x}&=1-F(F^{-1}(v))=1-v,\quad\mbox{for $\,v\in[0,1]$},
    \\\label{eq:F^-1hatpx}\actsymb[F^{-1}(v)]{\hat{p}}{x}&=1-\hat{F}(F^{-1}(v))=1-\frac{1}{\sum_{j=1}^Ns_j}\sum_{i=1}^Ns_i\mathbbm{1}_{\{U_i\leq v\}}.
\end{align}
We consider the time change $\,u=F(t)$, which tracks time based on how many people have passed away. Imagine an infinitely large pool where members follow the ideal survival rates. Then, any time point corresponds to a certain proportion of people who have passed away. In particular, the time change $F$ maps time $\,t\geq0\,$ to a number $\,u\in[0,1]$. Moreover, the maximal time $t$ that fulfils (\ref{eq:P>beta}) follows from the maximal $\,u=F(t)\,$ fulfilling
\begin{equation}\label{eq:Pu>beta}
    \mathbb{P}\Big[1+\varepsilon_2\geq\frac{\actsymb[F^{-1}(v)]{p}{x}}{\actsymb[F^{-1}(v)]{\hat{p}}{x}}\geq 1-\varepsilon_1\quad\mbox{for all $\,v\leq u$}\Big]\geq\beta.
\end{equation}
The maximal $u$ that fulfils (\ref{eq:Pu>beta}) is independent of the underlying mortality distribution because of (\ref{eq:F^-1px}) and (\ref{eq:F^-1hatpx}).

We need to identify a process within $\actsymb[F^{-1}(v)]{p}{x}/\actsymb[F^{-1}(v)]{\hat{p}}{x}$ that is asymptotically Gaussian to use results from probability theory. Indeed, a scaled version of $\,v\mapsto v-\hat{F}(F^{-1}(v))\,$ converges in distribution to a standard Brownian bridge; see Appendix \ref{ap-pro:Wealth-Donsker} for the proof. In particular, $\,v\mapsto(v-\hat{F}(F^{-1}(v)))/(1-v)\,$ is asymptotically a scaled and time-changed Brownian motion. The scale is $\,(\sum_{i=1}^Ns_i^2)^{1/2}/\sum_{i=1}^Ns_i$ and the time change is $\,v\mapsto v/(1-v)$.

We are ready to derive an approximate formula for the graphs in Figure \ref{fig:Poor-Rich}. Let $\,\varepsilon_2=\infty$, let $B$ be a standard Brownian motion and $\Phi$ the standard normal distribution function. The derivation is similar to the one in the appendix of \textcite{BeDo2021}. It approximates (\ref{eq:Pu>beta}) by interchanging the correct process with a Brownian motion and uses the reflection principle:
\begin{align*}
    \beta&\leq \mathbb{P}\Big[\,\frac{\actsymb[F^{-1}(v)]{p}{x}}{\actsymb[F^{-1}(v)]{\hat{p}}{x}}\geq 1-\varepsilon_1\quad\mbox{for all $\,v\leq u$}\Big]
    \\&=\;\mathbb{P}\Big[\frac{1-v}{1-\hat{F}(F^{-1}(v))}\geq 1-\varepsilon_1\quad\mbox{for all $\,v\leq u$}\Big]
    \\&=\mathbb{P}\Big[\frac{\varepsilon_1}{1-\varepsilon_1}\geq\frac{v-\hat{F}(F^{-1}(v))}{1-v}\quad\mbox{for all $\,v\leq u$}\Big]
    \\&\approx\mathbb{P}\Big[\frac{\varepsilon_1}{1-\varepsilon_1}\geq\sup_{v\leq u}\tfrac{\sqrt{\sum_{i=1}^Ns_i^2}}{\sum_{i=1}^Ns_i}B(\tfrac{v}{1-v})\Big]
    \\&=1-2\Phi\Big(-\frac{\varepsilon_1}{1-\varepsilon_1}\tfrac{\sum_{i=1}^Ns_i}{\sqrt{\sum_{i=1}^Ns_i^2}}\sqrt{\tfrac{1-u}{u}}\Big).
\end{align*}
Hence, the maximal time $t$ that fulfils (\ref{eq:P>beta}) for $\,\varepsilon_2=\infty\,$ follows from the approximation
\begin{equation}\label{eq:F(t)=u=frac}
    F(t)=u\approx\frac{1}{1+\tfrac{\sum_{i=1}^Ns_i^2}{(\sum_{i=1}^Ns_i)^2}\big(\tfrac{1-\varepsilon_1}{\varepsilon_1}\big)^2\big(\Phi^{-1}(\tfrac{1-\beta}{2})\big)^2}.
\end{equation}

Next, we numerically check the goodness of the above approximation. Figure \ref{fig:error} compares the approximation (\ref{eq:F(t)=u=frac}) with the values from the numerical analysis of Subsection \ref{subsection:problem}. 
\begin{center}
    \includegraphics[trim=0 10 0 30,width=0.45\linewidth]{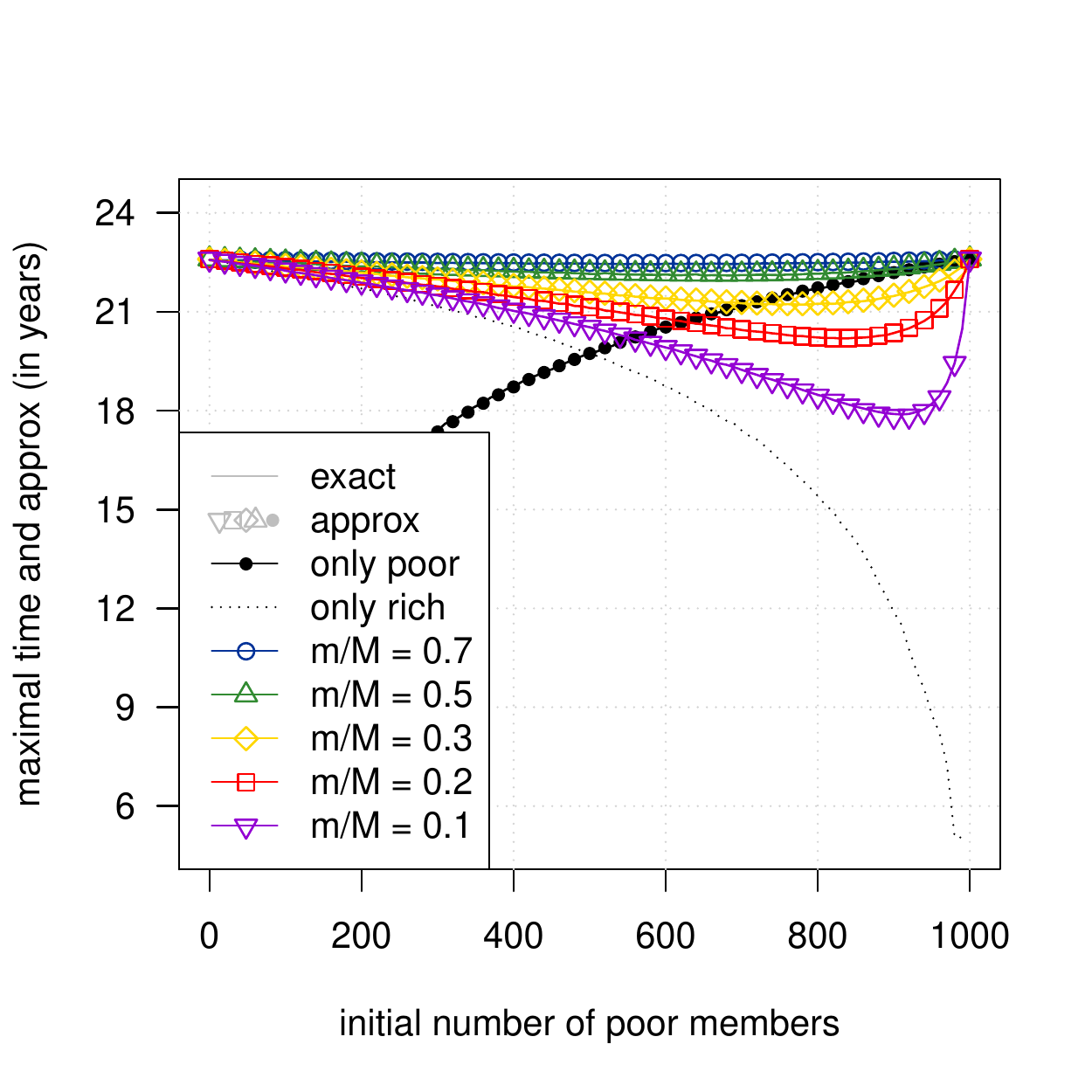}
    \includegraphics[trim=0 10 0 30,width=0.45\linewidth]{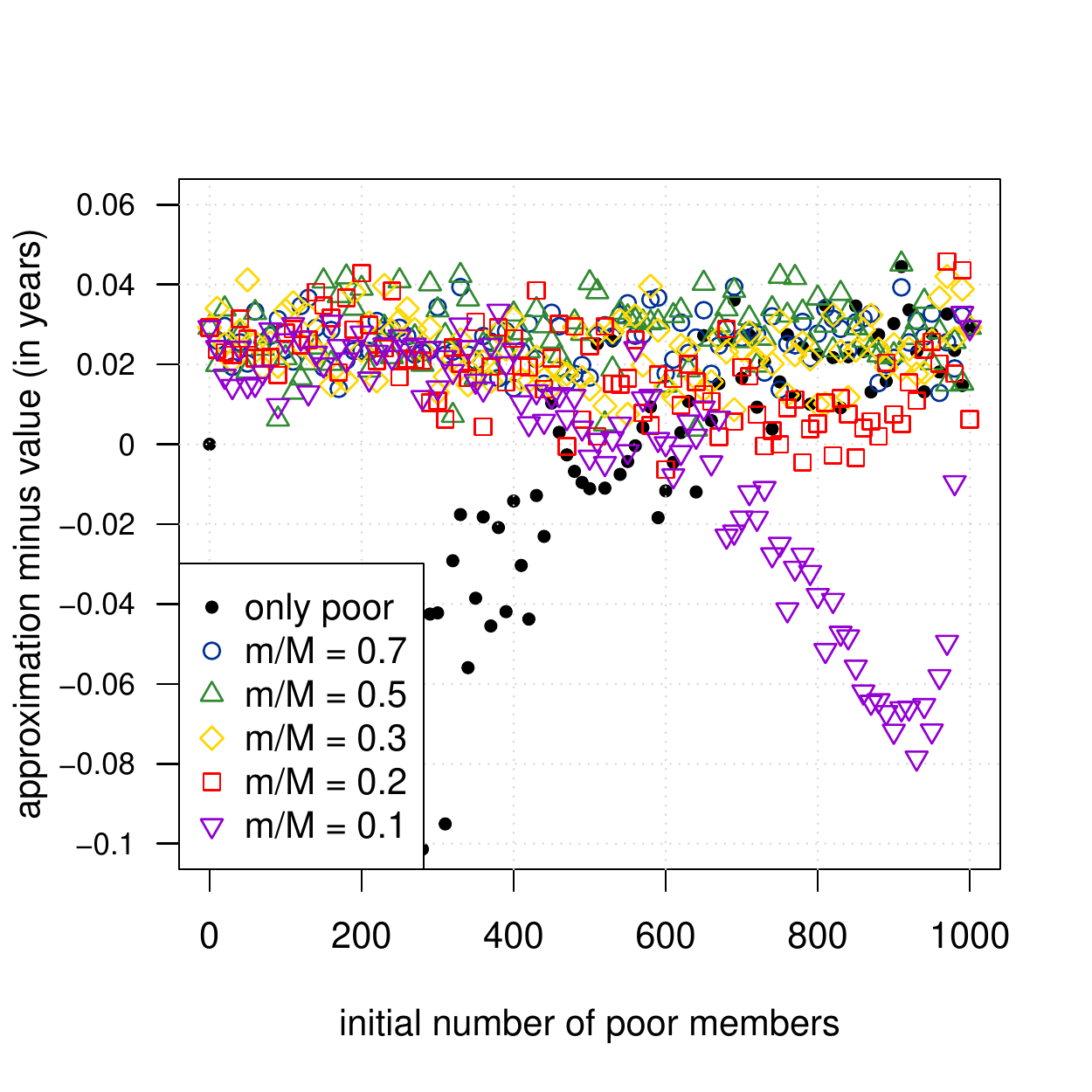}
    \captionof{figure}{The value and the approximation (left) and their difference (right) of the maximal time that a pooled annuity fund can provide a stable income with $\varepsilon_1=10\%$ and $\beta=90\%$. Pooling two subgroups with a total member size of 1000 (coloured symbols) and a single subgroup with varying member sizes (black dots). Different symbols indicate different income ratios (coloured symbols only). Members are initially 70 years old and follow UK period data for both sexes from the \textcite{HMD2020UK}.
    } 
    \label{fig:error}
\end{center}
There is no visible difference between the values from Subsection \ref{subsection:problem} and the approximation (\ref{eq:F(t)=u=frac}) in Figure \ref{fig:error} (left). We can find patterns only when we subtract the graphs in Figure \ref{fig:error} (left) and zoom in by a couple of magnitudes, see Figure \ref{fig:error} (right). The errors are more pronounced when the total number of members is low (solid black dots) or when there are only a few high-income members (reversed purple triangles). In both cases, only a few hundred members are the main driver of the income fluctuations; there are no other members in the first case, and only a few hundred members own the majority of the savings in the fund in the second case. Furthermore, looking at the y-axis of the difference in Figure \ref{fig:error} (right),  we can see that the error ranges from -0.08 to 0.06 years when there are at least a couple of hundred members, which yields an absolute error of at most a month. Overall, we can say that there is numerical evidence that approximation (\ref{eq:F(t)=u=frac}) works well for a pool in which more than a few hundred members are the main drivers of the income fluctuations based on the used mortality table.

It is worth noting that we can also study the error for the transformed time $u$ instead of the time $t$. The advantage is that that error in $u$ is independent of the underlying mortality distribution but loses its direct meaning in terms of time; we would need to look at the derivative of the mortality distribution function to link both errors. For the interested reader, the error for the transformed time $u$ ranges from -0.5\% to 0.2\%.

There is numerical evidence that the goodness of the approximation (\ref{eq:F(t)=u=frac}) is practically unaffected by $\,\varepsilon_1\in[0.05,0.1]\,$ and $\,\beta\in[0.90,0.99]$. \textcite{BeDo2021} reached that conclusion after a sensitivity analysis in the homogeneous case when all members have the same initial savings, i.e.\ when $1/N$ replaces $\,\sum_{i=1}^Ns_i^2/(\sum_{i=1}^Ns_i)^2$ in (\ref{eq:F(t)=u=frac}). We add to their sensitivity analysis in the direction of initial wealth distribution with the same conclusion.

Probably the most notable point of this subsection for our analysis is that the initial savings enter the approximation (\ref{eq:F(t)=u=frac}) only through the term $\,(\sum_{i=1}^Ns_i)^2/\sum_{i=1}^Ns_i^2$. We dub this term ``implied number of homogeneous members'' and study it in Section \ref{section:results}.

%------------------------
\subsection{Variance heuristic}
%------------------------

The main argument in the derivation in Subsection \ref{subsection:asymptotic} is that there is a process to which Donsker's theorem applies. Heuristically, we can imagine that there is such a process in most pooled annuity funds regardless of the specifics of the income and sharing rules because of the i.i.d.\ assumptions on the involved random variables. This rule of thumb can help us to identify the terms that link income fluctuation and initial savings.

Consider a retirement product that pools a group of same-aged people together and adjusts their income according to experienced mortality rates. Assume that the adjustments are consistent with the income computation, i.e.\ if there are infinitely many members, then every member would get a constant income (ignoring other risks but the idiosyncratic risk). Then, the income payments fluctuate around a baseline, and a sequence of i.i.d.\ random variables drive the income fluctuations. Due to Donsker's theorem, we assume that there is some underlying Brownian bridge that approximates the income fluctuations. In particular, the size of the income fluctuations depends only on a single parameter in front of the Brownian bridge. Due to the direct link between the Brownian bridge and the income fluctuations, we can find the parameter up to scaling by computing the variance of the first income payment.

\begin{ex}
    Let's consider the Annuity Overlay Fund by \textcite{DoGuNi2014}. Here, surviving members and also recently deceased members receive longevity credits. This change ensures actuarial fairness even for a heterogeneous composition of members with different mortality rates and initial savings. However, the Annuity Overlay Fund cannot provide a consistent income based on annuity prices because survivors do not receive all the funds. We need slightly higher prices than (\ref{eq:ddota}) to cover for the loss of funds that deceased members receive to ensure that an infinitely large pool of members yields a constant income for all members (ignoring other risks but the idiosyncratic risk), namely,
    \begin{equation}\label{eq:alpha}
        \alpha(x+t)=1+\sum_{\delta=1}^\infty(1+R)^{-\delta}\prod_{k=0}^{\delta-1}\frac{1}{2-\actsymb[1]{p}{x+t+k}}.
    \end{equation}
    If we use the above prices instead of $\ddot{a}(x+t)$ in (\ref{eq:C_i}) to calculate the income payments, then the income process fluctuates around its initial value in the Annuity Overlay Fund. In particular, we can apply our methodology to one cohort with the same mortality rates. Using the same notation for the individual income payments and fund accounts $C_i$ and $W_i$, but (\ref{eq:alpha}) instead of (\ref{eq:ddota}), and including the recently deceased members in the longevity credit calculation (\ref{eq:M_ioriginal}), we can show that
    \begin{equation*}
        C_i(1)=\frac{(W_i(0)-C_i(0))(1+R)}{\alpha(x+1)}\Big(1+\frac{\sum_{k=1}^N s_k\mathbbm{1}_{\{T_k\leq1\}}}{\sum_{k=1}^N s_k}\Big).
    \end{equation*}
    Computing the variance shows that the Annuity Overlay Fund depends on the same term, namely $\,(\sum_{i=1}^Ns_i)^2/\sum_{i=1}^Ns_i^2$, as the present pooled annuity fund. In particular, the actuarially fair Annuity Overlay Fund has the same issue with wealth heterogeneity as the pooled annuity fund in this paper. More precisely:
    \begin{align}
        \nonumber\mathrm{Var}\Big(C_i(1)\Big)
        &=\Big(\frac{(W_i(0)-C_i(0))(1+R)}{\alpha(x+1)}\Big)^2\frac{\sum_{k=1}^N s_k^2\mathrm{Var}(\mathbbm{1}_{\{T_k\leq1\}})}{(\sum_{k=1}^N s_k)^2}
        \\\nonumber&=\Big(\frac{(W_i(0)-C_i(0))(1+R)}{\alpha(x+1)}\Big)^2\frac{\sum_{k=1}^N s_k^2\actsymb[1]{p}{x}(1-\actsymb[1]{p}{x})}{(\sum_{k=1}^N s_k)^2}
        \\\label{eq:Var(C)}&=\actsymb[1]{p}{x}(1-\actsymb[1]{p}{x})\Big(\frac{(W_i(0)-C_i(0))(1+R)}{\alpha(x+1)}\Big)^2\frac{\sum_{k=1}^N s_k^2}{(\sum_{k=1}^N s_k)^2}.
    \end{align}
\end{ex}    

\begin{ex}
    It is worth mentioning how the approach to computing the variance applies in the situation of the present pooled annuity fund. There seems to be no explicit formula for the variance of the income payment because the random variable $\actsymb[t]{\hat{p}}{x}$ appears in the denominator of the income in (\ref{eq:C_i(t)=C_i(0)}). Instead, we look at the reciprocal $\,\actsymb[t]{\hat{p}}{x}/\actsymb[t]{p}{x}$, which is also constant if there are infinitely many members. Any upward (downward) excursion of the income payments is a downward (upward) excursion for the reciprocal and vice versa. Calculating the variance of $\,\actsymb[t]{\hat{p}}{x}/\actsymb[t]{p}{x}$ leads to the term $\,(\sum_{i=1}^Ns_i)^2/\sum_{i=1}^Ns_i^2$. More precisely:
    \begin{align}\label{eq:Var(hatp/p)}
        \mathrm{Var}\Big(\frac{\actsymb[1]{\hat{p}}{x}}{\actsymb[1]{p}{x}}\Big)&=\mathrm{Var}\Big(\frac{\sum_{i=1}^Ns_i\mathbbm{1}_{\{T_i>1\}}}{\actsymb[1]{p}{x}\sum_{i=1}^Ns_i}\Big)
        =\frac{\sum_{i=1}^N s_i^2\mathrm{Var}(\mathbbm{1}_{\{T_i>1\}})}{(\actsymb[1]{p}{x})^2(\sum_{k=1}^N s_i)^2}
        =\frac{1-\actsymb[1]{p}{x}}{\actsymb[1]{p}{x}}\frac{\sum_{i=1}^N s_i^2}{(\sum_{k=1}^N s_i)^2}.
    \end{align}
\end{ex}

%------------------------
\section{Implied number and beneficial groups}\label{section:results}
%------------------------

We identified the term $\,(\sum_{i=1}^Ns_i)^2/\sum_{i=1}^Ns_i^2\,$ as how the initial savings impact the stability of pooled annuity funds in Section \ref{section:derivation}. We will show how this term implies the observations from our numerical analysis in Subsection \ref{subsection:problem}. In particular, recall Remark \ref{rmk:graph}(a)--(e), which we will show in the following general sense:
\begin{enumerate}[label=(\alph*)]
    \item 
        wealth heterogeneity negatively affects the stability of income payments,
    \item
        richer members benefit from pooling their risk with poorer members,
    \item 
        poor members might prefer a smaller homogeneous than a larger heterogeneous pool,
    \item 
        if all savings amounts are at most 2 times the lowest savings amount, then all members benefit from pooling their funds together, and the income stability is close to the homogeneous case,
    \item
        if any savings amount is more than 2 times the lowest amount, then poor members tend to prefer no wealthy members unless there are comparable many wealthy members.
\end{enumerate}
Moreover, we will use the term $\,(\sum_{i=1}^Ns_i)^2/\sum_{i=1}^Ns_i^2\,$ to develop a criterion to determine whether or not a given group of people benefit in terms of income stability from being together. 

%------------------------
\subsection{The implied number of homogeneous members}\label{subsection:implied number}
%------------------------

We begin with some terminology and give the term $\,(\sum_{i=1}^Ns_i)^2/\sum_{i=1}^Ns_i^2\,$ a name, which is the basis of our mathematical analysis.
\begin{defi}\label{defi:impliednumber}
    The savings of a same-aged pool is a vector of strictly positive numbers $\,s=(s_i)_{i=1}^N$. We call
    \begin{equation*}
        \nu(s)=\frac{\big(\sum_{i=1}^Ns_i\big)^2}{\sum_{i=1}^Ns_i^2}
    \end{equation*}
    its implied number of homogeneous members.
\end{defi}

It is crucial to realise the simple relationship between the implied number of homogeneous members and the diversification of idiosyncratic risk: the higher the number, the more stable the income payments of the fund. We emphasise this relationship in the following proposition and take it as given in the rest of our analysis.
\begin{pro}
    Consider the savings of two same-aged pools $s$ and $t$ with $\,\nu(s)\leq\nu(t)$. Then 
    \begin{enumerate}[label=(\alph*)]
    \item 
        The approximation (\ref{eq:F(t)=u=frac}) of the maximal time that fulfils (\ref{eq:P>beta}) of Definition \ref{defi:P>beta} is longer for the savings $t$ than for the savings $s$.  
    \item 
        The variance (\ref{eq:Var(hatp/p)}) is smaller for savings $t$ than for savings $s$.
    \item           
        In the context of the Annuity Overlay Fund by \textcite{DoGuNi2014}, the income variance (\ref{eq:Var(C)}) of each member is smaller for savings $t$ than for savings $s$.
    \end{enumerate}
    \begin{proof}
        (\ref{eq:F(t)=u=frac}), (\ref{eq:Var(C)}) and (\ref{eq:Var(hatp/p)}) are monotone in the implied number of homogeneous members.
    \end{proof}
\end{pro}

The name ``implied number of homogeneous members'', for short ``implied number'', stems from searching for the size of a homogeneous pool of members that yields the same stability as the pool with initial savings $\,s=(s_i)_{i=1}^N$. Even though the implied number $\nu(s)$ is typically not an integer, we can still place the income stability between the two homogeneous pools with $\lfloor\nu(s)\rfloor$ and $\lceil\nu(s)\rceil$ members. For example, consider a pooled annuity fund with 1000 initial members, of whom 500 have savings of \pounds100k and 500 have savings of \pounds200k. Equation (\ref{eq:F(t)=u=frac}) implies that the income payments of such a pool stay above 90\% of their initial value in 90\% of all cases, i.e.\ $\,\varepsilon=0.1\,$ and $\,\beta=90\%$, for at least 22.19 years based on the used mortality table in Subsection \ref{subsection:problem}. Using 900 members with equal savings instead, (\ref{eq:F(t)=u=frac}) gives 22.19 years again. Even more, the maximal time that the heterogeneous 1000 member pool and the homogeneous 900 member pool provide a stable income coincides for all $\varepsilon$ and $\beta$. The reason for this equivalence is that equation (\ref{eq:F(t)=u=frac}) for a heterogeneous group differs from a homogeneous group only by replacing the implied number $\nu(s)$ with the total number of members $N$; and here, $\,900=(500\times100+500\times200)^2/(500\times100^2+500\times200^2)=\nu(s)$.

Moreover, the name ``implied number'' suggests the counting of members. Indeed, some counting laws hold in a weak sense. For example, adding a member increases the implied number by at most 1. In particular, the total number of members bounds the implied one. The reference to ``homogeneous members'' is suggestive as well. Fixing the total number of members, the implied one achieves its maximum in the homogeneous case only. And, some properties from the homogeneous case, like being independent of the scale of the savings amounts, carry over to the implied one. 

Because the total number of members bounds the implied one with equality only in the case of equal savings, we have that a homogeneous group yields the highest income stability for a given total member size. In particular, any deviation from the homogeneous case implies worse income stability. \textcite[Section 4.4]{FoSa2016} already found this in a one-period pooled annuity fund. Hence, wealth heterogeneity negatively affects the income stability in all pooled annuity funds, see Remark \ref{rmk:graph}(a). 

We summarise this result and the previously mentioned facts in the following lemma. Lemma \ref{lem:number}(a) shows that the total number of members bounds the implied one with equality only when the savings are equal. Lemma \ref{lem:number}(b) shows that adding a new member to the pool increases the number by at most 1. And \ref{lem:number}(c) proves that the implied number stays unchanged when we scale the savings.
\begin{lem}\label{lem:number}
    Consider savings $\,s=(s_i)_{i=1}^N$, its extension by one amount $\,s(x)=(s_i)_{i=1}^{N+1}$ where $\,s_{N+1}=x>0$, and its scale $\,\lambda s=(\lambda s_i)_{i=1}^N$ for $\,\lambda>0$. Then:
    \begin{enumerate}[label=(\alph*)]
        \item 
            $\nu(s)\leq N\,$ with equality only if $\,s_i=s_j\,$ for all $i$ and $j$,
        \item 
            $x\mapsto\nu(s(x))\,$ as a function on $(0,\infty)$ has a unique stationary point at $\,x^*=\sum_{i=1}^Ns_i^2/\sum_{i=1}^Ns_i$, at which it achieves its unique global maximum $\,\nu(s)+1$,
        \item
            $\nu(s)=\nu(\lambda s)$.
    \end{enumerate}
\begin{proof}
    (a) is a consequence of the Cauchy-Schwarz inequality applied to the two vectors $s$ and $(1)_{i=1}^N$. Equality can only appear if $s$ and $(1)_{i=1}^N$ are linear dependent, i.e.\ the entries of $s$ are equal. 
    
    (b) Let $\,b=\sum_{i=1}^Ns_i$, $\,c=\sum_{i=1}^Ns_i^2$. Note that $\,b,c>0$. Then,
    \begin{equation*}
        0=\partial_x\nu(s(x))=\partial_x\frac{(b+x)^2}{c+x^2}=\frac{2(b+x)}{(c+x^2)^2}\big[(c+x^2)-x(b+x)\big]\;\;\Rightarrow\;\;x=\frac{c}{b}.
    \end{equation*}
    Hence, $x^*=c/b\,$ is the unique stationary point of $\nu(s(x))$ in the positive half line. Moreover,
    \begin{align*}
        \nu(s(x^*))&=\frac{(b+\frac{c}{b})^2}{c+(\frac{c}{b})^2}=\frac{\frac{1}{b^2}(b^2+c)^2}{\frac{c}{b^2}(b^2+c)}=\frac{b^2+c}{c}=\nu(s)+1,
        \\\nu(s(x))&=\frac{(b+x)^2}{c+x^2}=\frac{(b/x+1)^2}{c/x^2+1}\xrightarrow[]{x\uparrow\infty}\frac{1^2}{1}=1,
        \\\nu(s(x))&\xrightarrow[]{x\downarrow0}\nu(s)>0.
    \end{align*}
    Overall, $\nu(s(x))$ is strictly larger at its unique stationary point $x^*$ than on its boundary. Hence, $x^*$ is the point of the unique global maximum.  
    
    (c) is a direct consequence of the definition because
    \begin{gather*}
        \nu(\lambda s)=\frac{\big(\sum_{i=1}^N\lambda s_i\big)^2}{\sum_{i=1}^N(\lambda s_i)^2}
        =\frac{\lambda^2\big(\sum_{i=1}^Ns_i\big)^2}{\lambda^2\sum_{i=1}^Ns_i^2}=\nu(s).
    \end{gather*}
\end{proof}
\end{lem}

Since income stability increases when the implied number increases, we can determine when one pool diversifies idiosyncratic risk better than another by comparing their numbers. For example, as in Subsection \ref{subsection:problem}, consider two groups with different savings. The members of the first group (poor) have fewer savings than the members of the second group (rich). We can show that the second group benefits from increased income stability by pooling their funds with the first group by comparing the implied numbers of the second group with the merged group. In particular, the following proposition shows that Remark \ref{rmk:graph}(b) holds in general.
\begin{pro}\label{pro:wealtherbetteroff}
    Consider the savings of two same-aged pools $\,s=(s_i)_{i=1}^N$ and $\,t=(t_i)_{i=1}^M$ with $\,\max\{s_i\}\leq\min\{t_i\}$. Let $\,u=(u_i)_{i=1}^{N+M}$ be the savings of the merged pool, i.e.\ $u_i=t_i\,$ for $\,i\leq N\,$ and $\,u_i=s_{i-N}$ for $\,i>N$. Then, $\,\nu(t)\leq\nu(u)$.
\begin{proof} 
    Let $\,t(x)=(t_i)_{i=1}^{M+1}\,$ with $\,t_{M+1}=x>0\,$ and assume that $\,x\leq\min\{t_i\,|\,1\leq i\leq M\}$. Then,
    \begin{align*}
        x\sum_{i=1}^Mt_i=\sum_{i=1}^Mxt_i\leq\sum_{i=1}^Mt_i^2\;\;\Rightarrow\;\;x\leq\frac{\sum_{i=1}^Mt_i^2}{\sum_{i=1}^Mt_i}=x^*.
    \end{align*}
    Lemma \ref{lem:number}(b) states that the above $x^*$ is the unique stationary point/point of global maximum of $\nu(t(x))$. Hence, $\nu(t(x))$ is increasing in $(0,x^*]$. Since $\,\nu(t(x))\rightarrow\nu(t)\,$ for $x\downarrow0$, it holds that $\,\nu(t(x))\geq\nu(t)$.
    
    Extending $t$ and reducing $s$ by the highest savings amount $x$ of $s$ repeatedly yields a monotone increasing chain of implied numbers of homogeneous members, which results in $\,\nu(u)\geq\nu(t)$.
\end{proof}
\end{pro}

We can prove Remark \ref{rmk:graph}(d) in the case of two homogeneous subgroups. However, we delegate this result to Subsection \ref{subsection:beneficial group}, where we introduce a concept which allows us to prove this result beyond the case of two homogeneous subgroups. Instead, we focus on the case when the highest savings amount is more than 2 times the lowest amount and establish Remark \ref{rmk:graph}(e).

Again, we consider two groups with different savings. The members of the first group (poor) have fewer savings than the members of the second group (rich). Unlike before, we change the wealth distribution by increasing the number of poor members. We compare the mixed pool with the pool that exclusively consists of poor members. Surprisingly, even though this increases the stability for the rich and the poor, the fund exclusively consisting of poor members yields a higher implied number with enough poor members than the mixed pool. In particular, when a fund attracts more and more members with low savings, the group without members that have more than double the savings amounts of the poor yields a higher implied number than the whole pool. Note that this gives us an easy way to construct examples described in Remark \ref{rmk:graph}(c). More precisely: 
\begin{pro}\label{pro:more-poor}
    Consider savings $\,s=(s_i)_{i=1}^N$ and $\,t=(t_i)_{i=1}^M$ with 
    \begin{equation*}
       2\,\frac{\sum_{i=1}^Ns_i^2}{\sum_{i=1}^Ns_i}<\frac{\sum_{i=1}^Mt_i^2}{\sum_{i=1}^Mt_i}.
    \end{equation*}
    For integer $\lambda>0$, let $u$ be the vector of savings of $t$ and $\lambda$ copies of $s$, and let $v$ be the savings of $\lambda$ copies of $s$. Then, there is  $\,\lambda^*\geq0\,$ such that for all $\lambda>\lambda^*$ the inequality $\,\nu(u)<\nu(v)\,$ holds true. 
\begin{proof}
    Let $\,\lambda>0$ be such that 
    \begin{align*}
        \frac{1}{\lambda}<\frac{\sum_{i=1}^Ns_i}{\sum_{i=1}^Mt_i}\frac{\sum_{i=1}^Ns_i}{\sum_{i=1}^Ns_i^2}\big(\frac{\sum_{i=1}^Mt_i^2}{\sum_{i=1}^Mt_i}-2\,\frac{\sum_{i=1}^Ns_i^2}{\sum_{i=1}^Ns_i}\big).
    \end{align*}
    Then some rearranging and some algebra yields
    \begin{align*}
        1+2\lambda\,\frac{\sum_{i=1}^Ns_i}{\sum_{i=1}^Mt_i}&<\lambda\,\frac{(\sum_{i=1}^Ns_i)^2}{\sum_{i=1}^Ns_i^2}\frac{\sum_{i=1}^Mt_i^2}{(\sum_{i=1}^Mt_i)^2}
        \\
        \big(\sum_{i=1}^Mt_i\big)^2+2\lambda\big(\sum_{i=1}^Ns_i\big)\big(\sum_{i=1}^Mt_i\big)&<\lambda\,\frac{(\sum_{i=1}^Ns_i)^2}{\sum_{i=1}^Ns_i^2}\sum_{i=1}^Mt_i^2
        \\
        \big(\sum_{i=1}^Mt_i\big)^2+2\lambda\big(\sum_{i=1}^Ns_i\big)\big(\sum_{i=1}^Mt_i\big)+\big(\lambda\sum_{i=1}^Ns_i\big)^2&<\lambda\,\frac{(\sum_{i=1}^Ns_i)^2}{\sum_{i=1}^Ns_i^2}\sum_{i=1}^Mt_i^2+\big(\lambda\sum_{i=1}^Ns_i\big)^2
        \\
        \big(\sum_{i=1}^Mt_i+\lambda\sum_{i=1}^Ns_i\big)^2&<\lambda\,\frac{(\sum_{i=1}^Ns_i)^2}{\sum_{i=1}^Ns_i^2}\big(\sum_{i=1}^Mt_i^2+\lambda\sum_{i=1}^Ns_i^2\big)
        \\
        \frac{(\sum_{i=1}^Mt_i+\lambda\sum_{i=1}^Ns_i)^2}{\sum_{i=1}^Mt_i^2+\lambda\sum_{i=1}^Ns_i^2}&<\lambda\,\frac{(\sum_{i=1}^Ns_i)^2}{\sum_{i=1}^Ns_i^2}
        \\
        \nu(u)&<\nu(v).
    \end{align*}
\end{proof}
\end{pro}

We know that equal savings yield the best possible income stability for a fixed number of members, i.e.\ the total number $N$ equals the implied one. Unfortunately, the implied one can take any value between 1 and $N$. A single person with disproportionate more funds than the rest reduces the implied number to its minimum at 1. We saw this in the proof of Lemma \ref{lem:number}(b), where we showed that $\,\nu(s(x))\rightarrow1\,$ for $\,x\uparrow\infty$. 

It is possible to impose income bounds to avoid situations where a single person owns disproportionately many funds. Unfortunately, an unfavourable wealth distribution could dramatically reduce the implied number even in such a case. We show in the following proposition that the worst possible wealth distribution is the one with one homogeneous poor subgroup plus one homogeneous rich subgroup (the situation of Subsection \ref{subsection:problem}). And, we find the worst possible implied number by optimising the setting of Subsection \ref{subsection:problem}.

The following proposition/remark is a warning of what could happen if we allow members to join a pooled annuity fund without restriction due to an unfavourable savings distribution. On the other hand, we will also use it to demonstrate that the impact of wealth heterogeneity is small when the income bounds are close to each other. We give examples after we have introduced the unintuitive formula.
\begin{pro}\label{pro:>N4mM/(m+M)}
    Consider the space $\mathcal{S}$ of savings $\,s=(s_i)_{i=1}^N$ with fixed number of members $N$ and fixed bounds $\,m\leq s_i\leq M\,$ for all $\,i\,$ where $\,M>m>0$. Then
    \begin{equation*}
        N\frac{4mM}{(m+M)^2}\leq\inf_{s\in\mathcal{S}}\nu(s)\leq N\frac{4mM}{(m+M)^2}(1+\frac{M^3}{4m^3N^2}).
    \end{equation*}
\begin{proof}
    The function $\,s\mapsto\nu(s)\,$ is continuous on the compact space $\,\mathcal{S}=[m,M]^N$, hence $\nu$ restricted to $\mathcal{S}$ achieves its infimum in $\mathcal{S}$. Let $s^*$ be the point where $\nu$ achieves its infimum. Then $s^*$ is also the point where $\nu(s)$ achieves its infimum when we only vary the $j$th coordinate $s_j\in[m,M]$ and fix $\,s_i=s_i^*\,$ for all $\,i\neq j$. Now, Lemma \ref{lem:number}(b) asserts that any extrema in $(m,M)$ is the unique global maximum in $[m,M]$, in particular, $s_j^*\notin(m,M)$. Thus, $s_j^*$ is $m$ or $M$ for all $j$. In particular, let $n$ be an integer, then
    \begin{equation}\label{eq:inf=Bin}
        \inf_{s\in\mathcal{S}}\nu(s)=\min_{0\leq n\leq N} N\frac{(M\frac{n}{N}+m(1-\frac{n}{N}))^2}{M^2\frac{n}{N}+m^2(1-\frac{n}{N})}.
    \end{equation}
    
    Next we optimize (\ref{eq:inf=Bin}) with respect to $n/N$. Consider the function $\,f(x)=(ax+b)^2/(cx+d)\,$ for $\,x\geq0$ with $\,a=M-m$, $\,b=m$, $\,c=M^2-m^2$, and $\,d=m^2$. Note that  $\,a,b,c,d>0\,$ and that $f$ is differentiable. The function $f$ has a unique stationary point at $\,x^*=(bc-2ad)/ac=m/(m+M)\in(0,1)$. The unique stationary point $x^*$ must be the point of the global minimum because $\,f(0)=f(1)=1<\infty=f(\infty)$. Thus
    \begin{equation*}
        \inf_{s\in\mathcal{S}}\nu(s)\geq N\frac{(Mx^*+m(1-x^*))^2}{M^2x^*+m^2(1-x^*)}=N\frac{4mM}{(m+M)^2}.
    \end{equation*}
    This completes the proof of the lower bound of the statement. We redirect the interested reader for the second inequality to \ref{pro:>N4mM/(m+M)Part2} in the Appendix
\end{proof}
\end{pro}

\begin{rmk} 
    The lower bound of Proposition \ref{pro:>N4mM/(m+M)} holds with equality when $\,Nm/(m+M)\,$ is an integer. 

    We can rewrite the problem to find the worst possible homogeneous number to find the worst possible wealth distribution. However, the fixed $N$ implies a restriction on the possible wealth distributions. Taking $\,N\uparrow\infty\,$ removes this restriction. In particular, the difference in the bounds of Proposition \ref{pro:>N4mM/(m+M)} is a discretisation error. More precisely,
    \begin{equation*}
        \lim_{N\rightarrow\infty}\inf_{s\in\mathcal{S}}\frac{\nu(s)}{N}=\inf_{m\leq X\leq M}\frac{\mathbb{E}^2[X]}{\mathbb{E}[X^2]}=\frac{4mM}{(m+M)^2},
    \end{equation*}
    in which $X$ is any random variable that fulfils $\,m\leq X\leq M$.
\end{rmk}

The last proposition/remark quantifies the possible negative impact of allowing members freely to a pool. For example, take a homogeneous group where each member has \pounds100k. Assume that the pool attracts some new members with \pounds1000k. Then, their implied number could drop to $\,33\%\approx40/121=4\times100\times1000/(100+1000)^2\,$ of the total member size. More specifically, a homogeneous group with 1000 members attracting 100 wealthy retirees could see a drop of the implied number from 1000 to $\,364\approx1100\times40/121$, i.e.\ allowing members freely to the pool may effectively cut the size of the pool. 

Proposition \ref{pro:>N4mM/(m+M)} explains why wealth heterogeneity barely affects the stability of income payments when the highest savings amount is at most 2 times the lowest amount, i.e.\ Remark \ref{rmk:graph}(d). The implied number can only drop to $\,89\%\approx8/9=4\times1\times2/(1+2)^2\,$ of the best possible situation.

%------------------------
\subsection{Beneficial groups}\label{subsection:beneficial group}
%------------------------

Next, we introduce our tool to answer whether or not all members benefit from pooling their funds together. 

We looked at a group of members split into two subgroups (poor and rich) in Section \ref{subsection:problem} and compared the income stability of three pooled annuity funds: 1st the fund of poor members, 2nd the fund of rich members, and 3rd the fund of all members. However, there is no reason to restrict us to those three groups. There could be a subgroup comprised of some poor and some rich with higher income stability in the fund than any of the three previous funds. If such a subgroup exists, some members of the whole group are at a disadvantage for being in the pool because they could increase their income stability by leaving the pool and forming a new fund. We only consider a fund to be ``beneficial'' for all members if there is no better subgroup than the whole group. Using the connection between the implied number and income stability (diversifying idiosyncratic risk), we say that a group is ``beneficial'' if the whole group maximises the implied number under all subgroups, which leads to the following definition.
\begin{defi}\label{defi:beneficial} 
    We call the savings $\,s=(s_i)_{i=1}^N$ of a same-aged pool beneficial if 
    \begin{equation*}
        \nu(s)=\max_{n_i\in\{0,1\}}\frac{\big(\sum_{i=1}^Ns_in_i\big)^2}{\sum_{i=1}^Ns_i^2n_i}
    \end{equation*}
    with the convention that $\,0/0=0$.
\end{defi}

In Remark \ref{rmk:graph}(d), we observed that the income stability is better in a larger mixed pool than in a smaller homogeneous pool when the highest savings amount is at most 2 times the smallest amount. Indeed, the income stability is at least as good in a larger mixed pool than in any strict subgroup when the highest saving is at most 2 times the lowest amount. More precisely, the following proposition shows that any group with such a savings distribution is beneficial. 
\begin{pro}\label{pro:double}
    Any savings $(s_i)_{i=1}^N$ such that $\,\frac{\max\{s_i\}}{\min\{s_i\}}\leq2$ is beneficial.
\begin{proof}
    We prove the statement by contradiction and assume that a strict subgroup of savings achieves the maximum in Definition \ref{defi:beneficial}. After reindexing the savings amounts, we assume without loss of generality that there is $\,k<N\,$ such that the following holds.
    \begin{equation}\label{eq:k=max}
        \nu((s_i)_{i=1}^k)=\max_{n_i\in\{0,1\}}\frac{\big(\sum_{i=1}^Ns_in_i\big)^2}{\sum_{i=1}^Ns_i^2n_i}.
    \end{equation}
    Because of \ref{lem:number}(c), we can scale $\,(s_i)_{i=1}^N$ with $1/\min\{s_i\}$ and assume without loss of generality that $\,s_i\in[1,2]\,$ for all $i$. Hence, $\,s_{k+1}\leq2\,$ and $\,\sum_{i=1}^ks_i\leq\sum_{i=1}^ks_i^2$. Thus,
    \begin{align}
        \nonumber\frac{s_{k+1}}{\sum_{i=1}^ks_i^2}&\leq\frac{2}{\sum_{i=1}^ks_i}
        \\\nonumber1+\frac{s_{k+1}^2}{\sum_{i=1}^ks_i^2}&\leq1+\frac{2s_{k+1}}{\sum_{i=1}^ks_i}
        \\\nonumber1+\frac{s_{k+1}^2}{\sum_{i=1}^ks_i^2}&<\big(1+\frac{s_{k+1}}{\sum_{i=1}^ks_i}\big)^2
         \\\nonumber\frac{\sum_{i=1}^{k+1}s_i^2}{\sum_{i=1}^ks_i^2}&<\big(\frac{\sum_{i=1}^{k+1}s_i}{\sum_{i=1}^ks_i}\big)^2
        \\\nonumber\frac{\big(\sum_{i=1}^ks_i\big)^2}{\sum_{i=1}^ks_i^2}&<\frac{\big(\sum_{i=1}^{k+1}s_i\big)^2}{\sum_{i=1}^{k+1}s_i^2}
        \\\label{eq:double-nu-chain}\nu((s_i)_{i=1}^k)&<\nu((s_i)_{i=1}^{k+1}).
    \end{align}
    However, (\ref{eq:k=max}) contradicts (\ref{eq:double-nu-chain}). So, our original assumption is wrong, and the whole group achieves the maximum in Definition \ref{defi:beneficial}, implying that $(s_i)_{i=1}^N$ is beneficial.
\end{proof}
\end{pro}

A typical savings distribution of retirees might be more spread out than ``highest savings amount is at most 2 times the lowest amount'', even though this is a simple criterion to ensure that every member benefits from pooling their funds together. It comes down to whether or not we can determine if a given group is beneficial.

There is no standard way to maximise the implied number of a given group. For example, consider a pool with 1000 members. Then there are  $\,2^{1000}\approx10^{301}$ possible subgroups. Thus, a brute-force search would require evaluating around $\,10^{301}\,$ implied numbers, which is unmanageable with current computer power in a reasonable time. Similarly, attempting first-order-conditions with $\,0\leq n_i\leq1\,$ instead of $\,n_i\in\{0,1\}\,$ fail because the maximum is always on the boundary of its domain. However, it is intuitively clear that we only need to check a small class of subgroups. For example, we only considered three specific subgroups in Subsection \ref{subsection:problem} even though there are $\,2^{1000}$. Indeed, we can reduce the number of subgroups that we need to check to at most $N$ candidates when there are $N$ members, which is computational as good as it gets because the problem consists of at least considering the savings amounts of $N$ members.

To define the possible subgroups that potentially achieve the maximal implied number, we introduce another way to describe the savings of a group. So far, we have just listed the savings by individuals. Now, we consider the distinctively different savings amounts in increasing order and note how many members have a particular savings amount. That is called a hash map of savings in computer science. As it turns out, we only need to check the cumulative unions of subgroups with increasingly higher savings to determine the one with the maximal implied number. For example, consider a group where each member has \pounds100k, \pounds200k, \pounds300k, or \pounds400k savings. Then, we only need to check four subgroups to determine the one with the maximal implied number. The potential subgroups consist of 1st members with at most \pounds100k, 2nd members with at most \pounds200k, 3rd members with at most \pounds300k, and 4th all members.

In the following, we prove the cumulative unions of subgroups with increasingly higher savings contain all subgroups that achieve the maximal implied number. We begin by defining the hash map of savings amounts, i.e.\ listing the savings in increasing order and noting the number of members with each particular savings amount.
\begin{defi}\label{defi:sameagedsubgroup} 
    A hash map of savings of a same-aged pool is a pair of same length vectors $\,(z_i)_{i=1}^I,(N_i)_{i=1}^I$ of strictly positive numbers with $\,z_j\geq z_i$ for $\,j\geq i$. Its implied number of homogeneous members is
    \begin{equation*}
        \nu((z_i)_{i=1}^I,(N_i)_{i=1}^I)=\frac{\big(\sum_{i=1}^Iz_iN_i\big)^2}{\sum_{i=1}^Iz_i^2N_i}.
    \end{equation*}
    We call it beneficial if
    \begin{equation*}
        \nu((z_i)_{i=1}^I,(N_i)_{i=1}^I)=\sup_{0\leq n_i\leq N_i}\frac{\big(\sum_{i=1}^Iz_in_i\big)^2}{\sum_{i=1}^Iz_i^2n_i}
    \end{equation*}
    with the convention that $\,0/0=0$.
\end{defi}

Before we continue, it is worth noting that we allow the number of members $N_i$ with savings $z_i$ to be a non-integer. That helps to circumvent technical issues in the following proofs. For example, consider merging similar beneficial groups. We expect the result to be beneficial again. Indeed, the following proposition shows that, but more importantly, note how short the proof is.
\begin{pro}\label{pro:scale}
    Let $\,(z_i)_{i=1}^I,(N_i)_{i=1}^I$ be a hash map of savings. Let $\,\lambda>0$. Then, the hash map of savings $\,(z_i)_{i=1}^I,(\lambda N_i)_{i=1}^I$ is beneficial too.
\begin{proof}
    \begin{align*}
        \sup_{0\leq n_i\leq \lambda N_i}\frac{\big(\sum_{i=1}^Iz_in_i\big)^2}{\sum_{i=1}^Iz_i^2n_i}
        &=\sup_{0\leq n_i/\lambda\leq N_i}\frac{\big(\sum_{i=1}^Iz_in_i\big)^2}{\sum_{i=1}^Iz_i^2n_i}
        =\sup_{0\leq m_i\leq N_i}\frac{\big(\sum_{i=1}^Iz_im_i\lambda\big)^2}{\sum_{i=1}^Iz_i^2m_i\lambda}
        \\&=\lambda\sup_{0\leq m_i\leq N_i}\frac{\big(\sum_{i=1}^Iz_im_i\big)^2}{\sum_{i=1}^Iz_i^2m_i}=\lambda\,\nu((z_i)_{i=1}^I,(N_i)_{i=1}^I)
        \\&=\frac{\lambda^2\big(\sum_{i=1}^Iz_iN_i\big)^2}{\lambda\sum_{i=1}^Iz_i^2N_i}=\nu((z_i)_{i=1}^I,(\lambda N_i)_{i=1}^I).
    \end{align*}
\end{proof}
\end{pro}

We are ready to prove our main mathematical contribution. The following theorem shows that the cumulative unions of subgroups with increasingly higher savings contain all subgroups that achieve the maximal implied number of a group.
\begin{thm}\label{thm:optimal}
    Let $\,(z_i)_{i=1}^I,(N_i)_{i=1}^I$ be a hash map of savings. Then, there is at least one solution to the following optimisation problem with convention $\,0/0=0$,
    \begin{equation*}
        \sup_{0\leq n_i\leq N_i}\frac{\big(\sum_{i=1}^Iz_in_i\big)^2}{\sum_{i=1}^Iz_i^2n_i}.
    \end{equation*}
    And, any solution $(n^*_i)_{i=1}^I$ has an index $i^*$ such that 
    \begin{equation*}
        n^*_i=
        \begin{cases}
            N_i&\mbox{if}\quad i\leq i^*
            \\\hfill0&\mbox{if}\quad i>i^*
        \end{cases}.
    \end{equation*}
\begin{proof}
    For $\,\sum_{j=1}^In_i>0$, the Cauchy-Schwarz inequality implies that
    \begin{equation*}
        \frac{\big(\sum_{i=1}^Iz_in_i\big)^2}{\sum_{i=1}^Iz_i^2n_i}
        =\frac{\big(\sum_{i=1}^Iz_in_i\big/\sum_{j=1}^In_j\big)^2}{\sum_{i=1}^Iz_i^2n_i\big/\sum_{j=1}^In_j}\sum_{j=1}^In_j
        \leq\sum_{i=1}^In_i
        \xrightarrow{(n_i)\downarrow0}0.
    \end{equation*}
    Hence, the limit of the function $\,(n_i)_{i=1}^I\mapsto\sum_{i=1}^Iz_i^2n_i\big/\big(\sum_{i=1}^Iz_in_i\big)^2$ coincides with the convention that $\,0/0=0$. So, the function is continuous on its compact domain $\,\{(n_i)\,|\,0\leq n_i\leq N_i\;\;\mbox{for all $i$}\}\,$ and therefore achieves its supremum. Thus, the optimisation problem has a solution.
    
    Let $(n^*_i)_{i=1}^I$ be a solution to the optimisation problem and let $i^*$ be the index of its last non-zero element. We are going to show that $\,n^*_i<N_i\,$ for some $\,i\leq i^*$ leads to a contradiction.
    
    Assume that there is $\,n^*_j<N_j\,$ for some $j\leq i^*$. The optimality of the solution implies
    \begin{equation*}
        \frac{\big(\sum_{i\neq i^*}z_in^*_i\big)^2}{\sum_{i\neq i^*}z_i^2n^*_i}
        \leq\frac{\big(\sum_{i=1}^Iz_in^*_i\big)^2}{\sum_{i=1}^Iz_i^2n^*_i}
        \geq\frac{\big(z_j(N_j-n_j^*)+\sum_{i=1}^Iz_in^*_i\big)^2}{z_j^2(N_j-n_j^*)+\sum_{i=1}^Iz_i^2n^*_i}.
    \end{equation*}
    For ease of notation, let $\,a_1=z_{i^*}$, $a_2=z_j$, $M=n^*_{i^*}$, $N=N_j-n^*_j$, $b=\sum_{i\neq i^*}^Iz_in^*_i$, $c=\sum_{i\neq i^*}^Iz_i^2n^*_i$, which yields $\,M,N,a_1,a_2>0\,$ and $\,b,c\geq0\,$ and $\,a_1\geq a_2\,$ and 
    \begin{equation}\label{eq:double-inequality}
        \frac{b^2}{c}
        \leq\frac{(a_1M+b)^2}{a_1^2M+c}
        \geq\frac{(a_1M+a_2N+b)^2}{a_1^2M+a_2^2N+c}.
    \end{equation}
    Let $\,f(x)=(a_1M+a_2x+b)^2/(a_1^2M+a_2^2x+c)$. The function $f$ can only have one stationary point in the positive half line at $\,x^*=(a_1a_2M+a_2b-2a_1^2M-2c)/a_2^2$. Assume $x^*$ is negative. Then $f'$ cannot vanish in the positive half line. In particular, $f$ is strictly monotone in the positive half line (by continuity arguments applied to $f'$). As $\,f(x)>0\,$ for $\,x>0$, but $\,f(x)\rightarrow\infty\,$ for $\,x\uparrow\infty$, the function $f$ can only be strictly monotone when it increases in the positive half line. Overall, $f$ increases in the positive half line if $x^*$ is negative. However, that violates the second inequality in (\ref{eq:double-inequality}). Hence, $x^*\geq0$. That together with $\,a_1\geq a_2>0\,$ and $\,M,b\geq0\,$ yields
    \begin{gather}
        x^*=(a_1a_2M+a_2b-2a_1^2M-2c)/a_2^2\geq0,\nonumber 
        \\a_1^2M+a_1b-2a_1^2M-2c\geq0,\nonumber 
        \\a_1b\geq a_1^2M+2c\label{eq:2nd-inequality}
        .
    \end{gather}
    On the other side, rearranging the first inequality of (\ref{eq:double-inequality}) and using that $\,M,a_1>0\,$ and $\,c\geq0\,$ ($c=0\,$ implies $\,b=0$) yields
    \begin{gather}
        \frac{b^2}{c}
        \leq\frac{(a_1M+b)^2}{a_1^2M+c},\nonumber 
        \\(a_1^2M+c)b^2\leq c(a_1M+b)^2,\nonumber 
        \\a_1^2Mb^2+cb^2\leq ca_1^2M^2+2cba_1M+cb^2,\nonumber
        \\a_1b^2\leq ca_1M+2cb\label{eq:1st-inequality}
        .
    \end{gather}
    The two inequalities (\ref{eq:2nd-inequality}) and (\ref{eq:1st-inequality}) contradict each other. To see this, we put (\ref{eq:2nd-inequality}) into (\ref{eq:1st-inequality}) and compare it with (\ref{eq:2nd-inequality}) again. Note again that $\,M,a_1>0\,$ and that $\,b\geq0$,
    \begin{gather*}
        a_1^2Mb+2cb=(a_1^2M+2c)b\stackrel{(\ref{eq:2nd-inequality})}{\leq}(a_1b)b=a_1b^2\stackrel{(\ref{eq:1st-inequality})}{\leq}ca_1M+2cb,\nonumber
        \\a_1b\leq c
        .
    \end{gather*}
    Applying that to (\ref{eq:2nd-inequality}) and noting that $\,a_1^2M>0\,$ and $\,c\geq0\,$ implies $\,a_1b\geq a_1^2M+2c>c\geq a_1b$, hence $\,a_1b>a_1b$, a contradiction. 
    
    Overall, the original assumption is wrong and $\,n^*_j=N_j\,$ for all $\,j\leq i^*$.  
\end{proof}
\end{thm}

\begin{rmk}
    Theorem \ref{thm:optimal} is also true in the case when $(N_i)_{i=1}^I$ is weighted by some strictly positive constants $(\omega_i)_{i=1}^I$, i.e.\ we would be interested in \begin{align*}
         \nu((z_i)_{i=1}^I,(N_i)_{i=1}^I,(\omega_i)_{i=1}^I)=\frac{\big(\sum_{i=1}^Iz_iN_i\omega_i\big)^2}{\sum_{i=1}^Iz_i^2N_i\omega_i}.
    \end{align*}
    There is some indication that such a statistic is an appropriate extension of the implied number to account for mortality heterogeneity, i.e.\ pooled annuity funds with multiple cohorts, see \textcite[Eq.24]{DoGuNi2014} or \textcite[Eq.13]{FoSa2016} or \textcite{Wright2018}. Most notably, Theorem \ref{thm:optimal} might come in handy in future research.
\end{rmk}

We finish this section with a practical consideration about limiting the contributions to pooled annuity funds. If we restrict ourselves to beneficial groups, we make pooled annuity funds unattainable for people with high enough savings. However, we could introduce a savings amount that those people are allowed to contribute to the pool. For example, contributing the highest savings amount of a beneficial group guarantees that the fund stays beneficial. More precisely: 
\begin{pro}
    Consider beneficial savings $\,s=(s_i)_{i=1}^N$ and integer $\,M>N$. Then $\,t=(t_i)_{i=1}^M$ with $\,t_i=s_i$ for all $\,i\leq N\,$ and $\,t_i=\max\{s_i\}\,$ for all $\,i>N\,$ is beneficial.
\begin{proof}
    All solutions that achieve the maximum in Definition \ref{defi:beneficial} are under the cumulative union of subgroups with increasingly higher savings by Theorem \ref{thm:optimal}. Note that $s$ contains an entry that equals $\max\{s_i\}$ but not all the entries that equal $\max\{s_i\}$ in $t$ since $\,M>N$. In particular, $s$ cannot achieve the maximum for $t$. 
    
    Also, any cumulative union of subgroups with increasingly higher savings amounts of $s$ cannot achieve the maximum for $t$. If a subgroup would, then $s$ would be as good as that subgroup because $s$ is beneficial. However, that contradicts that $s$ cannot achieve the maximum for $t$.
    
    However, the only cumulative union of subgroups with increasingly higher savings of $t$ that is not under the cumulative unions of subgroups with increasingly higher savings of $s$ is $t$ itself. Hence, $t$ achieves the maximum in Definition \ref{defi:beneficial} by Theorem \ref{thm:optimal}, implying that $t$ is beneficial.
\end{proof}
\end{pro}

%---------------------------------------------
\subsection{A hypothetical application}
%---------------------------------------------

Theorem \ref{thm:optimal} suggests restricting participation to a pooled annuity fund based on income brackets to minimise income fluctuation, which stems from the wealth heterogeneity of its members. For example, consider an insurance company that helps employers to sponsor their employees' retirement plans. The employers want to build a pooled annuity fund. The employers know about the savings amounts of their employees in their defined contribution pension schemes. The employers are sure that most of their employees would participate in the pooled annuity fund because they would encourage it. Figure \ref{fig:Company} (left) shows the histogram of the distribution of savings from the 1000 employees who retire next year. Figure \ref{fig:Company} (right) is the corresponding analysis of the insurance company.
\begin{center}
    \includegraphics[trim=0 10 0 30,width=0.45\linewidth]{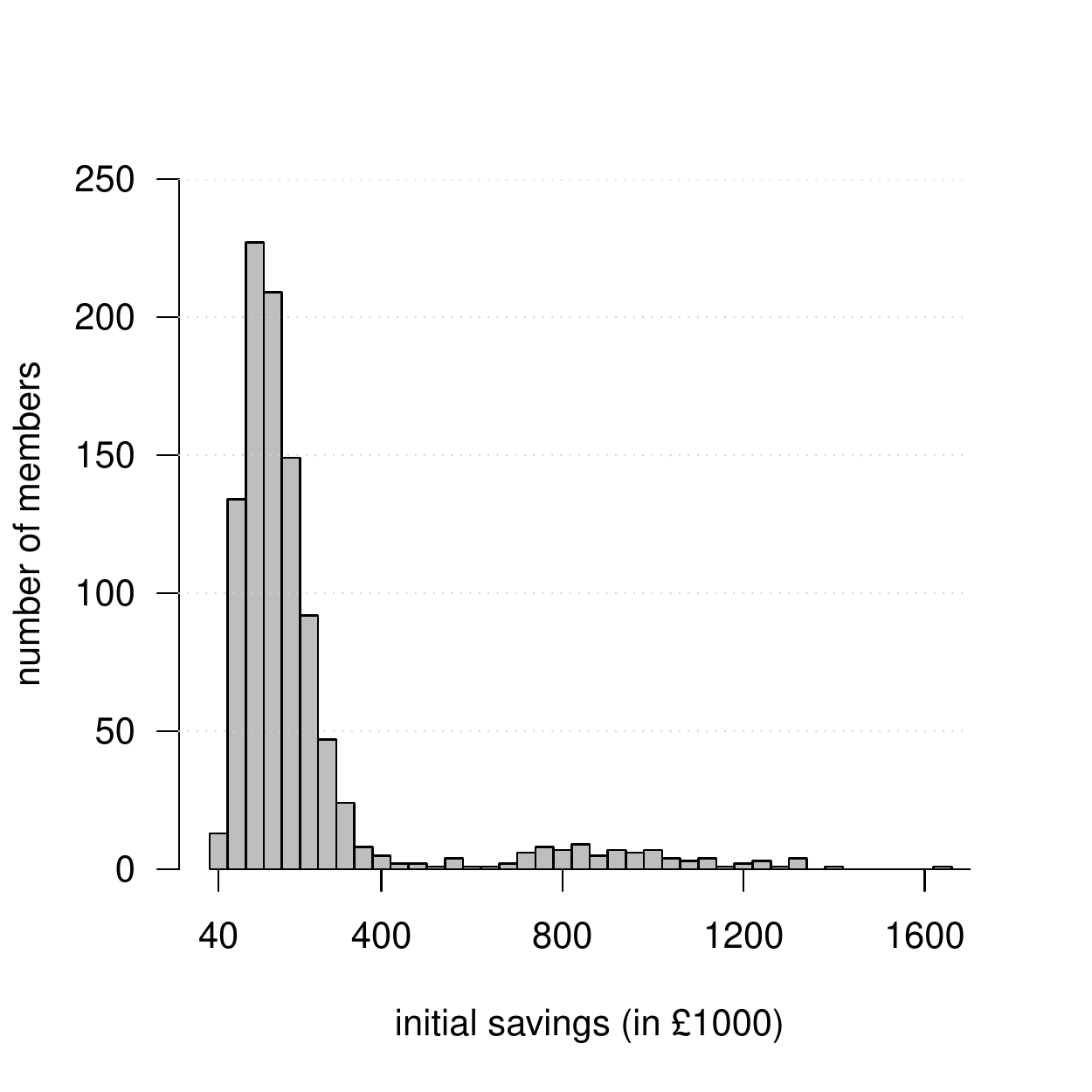}
    \includegraphics[trim=0 10 0 30,width=0.45\linewidth]{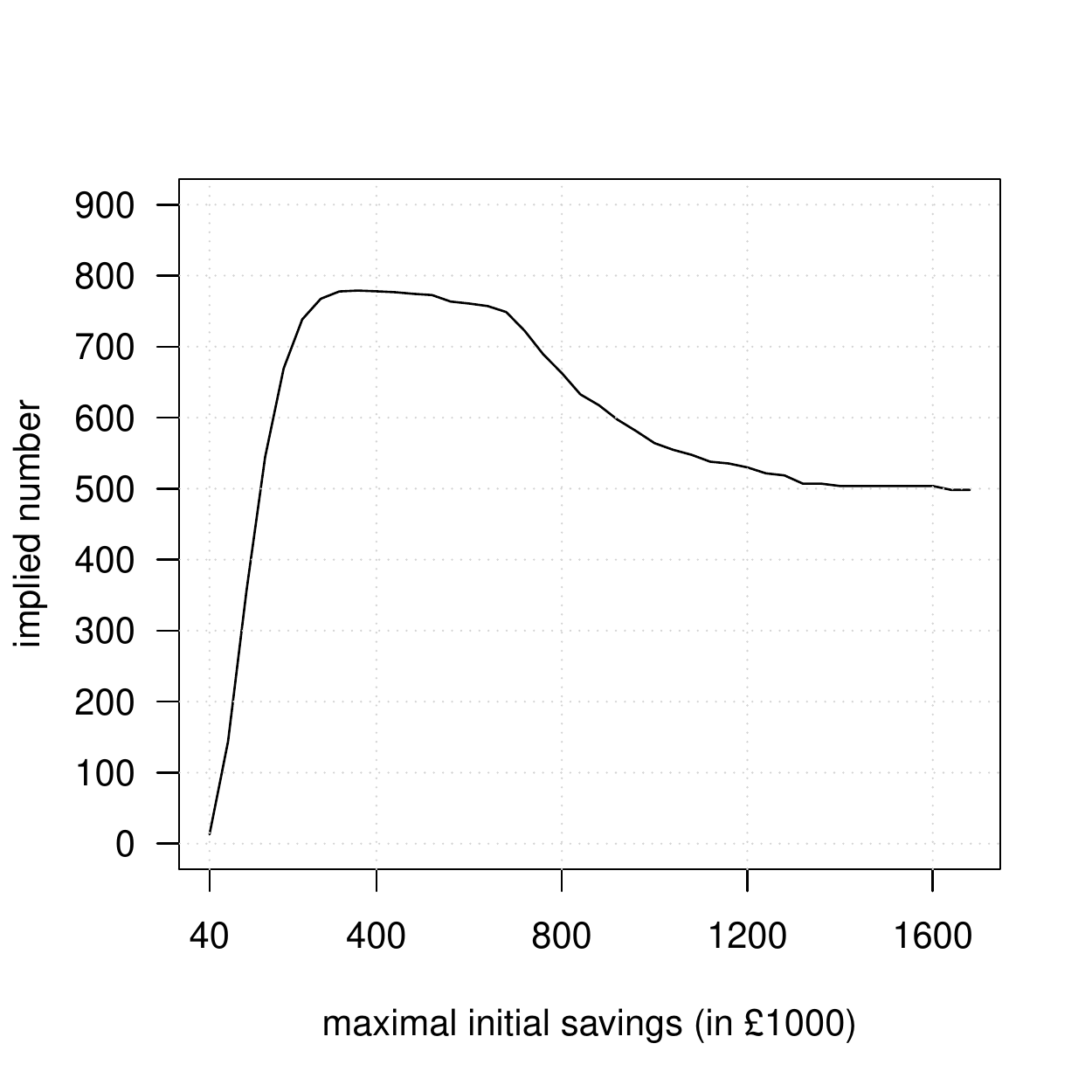}
    \captionof{figure}{A random sample of 1000 members split into $50$ subgroups (left) and the implied numbers of cumulative subgroups with increasingly higher savings (right). 
    %The difference of initial savings between two consecutive subgroups is $\pounds40\,000$.
    }
    \label{fig:Company} 
\end{center}

The insurance company suggests limiting the pooled annuity fund at a maximal savings level of \pounds360k to \pounds680k with 903 to 921 employees. The implied number varies insignificantly within that range from a minimum of about 757 and a maximum of about 779. In that range, members have at most a negligible interest in changing the pool.  

The insurance company warns the employers about allowing participation for all employees with their whole savings. An all-employee fund yields an implied number of about 498, i.e.\ a cut of $\,34\%\approx(757-498)/757\,$ to $\,36\%\approx(779-498)/779\,$ in effective participation. This disadvantages employees with low savings amounts in favour of a few employees with high savings amounts. The consequence could be that the pool could only deliver a stable income for 19.7 years instead of 21.5 to 21.6 years, which means a reduction of two years. The last statement is based (\ref{eq:F(t)=u=frac}) with the same mortality distribution as in Section \ref{subsection:problem} and ignoring systematic risks.

The insurance company encourages the employers to allow the wealthier employees with more than the maximal savings level to participate in the pooled fund by contributing the maximal savings amount instead of their whole savings. In this case, all members benefit from pooling their funds together regardless of how many wealthier employees participate.

%----------------------------------------------
\section{Conclusion}
%----------------------------------------------

We have investigated the impact of different initial savings amounts on the idiosyncratic risk of the pooled annuity fund by \textcite{PiVaDe2005} for one cohort. We have quantified the risk in terms of the time the pooled annuity fund provides a stable income to its members, which goes back to \textcite{BeDo2021}. We have found cases where pooling funds is not in the best interest of the whole group.

We have identified a parameter encapsulating the wealth heterogeneity of the pooled annuity fund and dubbed it the ``implied number of homogeneous members''. The name refers to the number of members an imaginary homogenous pool would have to achieve the same stability as the given one. It links directly to the time the fund can provide a stable income and the variance of the income payments. It also appears in other tontines like the actuarially fair fund by \textcite{DoGuNi2014}, the Annuity Overlay Fund. In particular, actuarial fairness does not prevent any issue steaming from the implied number or wealth heterogeneity.

We have focused our analysis on understanding how the implied number affects the stability of income payments. It has revealed that:
\begin{enumerate}[label=(\alph*)]
    \item[(i)] 
        wealth heterogeneity always negatively affects the stability of the fund (Lemma \ref{lem:number}(a)),
    \item[(ii)] 
        wealthy members benefit from pooling their funds with poorer members (Proposition \ref{pro:wealtherbetteroff}),
    \item[(iii)]
        if savings are at most a factor of 2 apart, all members benefit from pooling their funds together, and the stability is close to the homogeneous case (Proposition \ref{pro:>N4mM/(m+M)} and \ref{pro:double}),
    \item[(iv)]    
        if savings are more than a factor of 2 apart, then many poor members tend to prefer no wealthy members unless there are comparable many wealthy members (Proposition \ref{pro:more-poor}).
\end{enumerate}
In general, poorer members smooth the income payments of wealthy members, while wealthy members are the leading cause of large income fluctuations. The situation becomes more nuanced when there is a range of savings amounts, and single members are both: smoothers and the leading cause of large income fluctuations.
 
We have deemed a pool beneficial if the whole fund maximises the implied number under all subsets of members. Here, every member benefits from pooling their funds in terms of stability and nobody wants to change. In particular, if a pool is not beneficial, it is not in the best interest of all members to pool their funds together.

We can find all beneficial subgroups by starting from the poorest member and including members in increasing order of their savings, which is our main mathematical contribution (Theorem \ref{thm:optimal}). In particular, there is an efficient way to find beneficial groups. We imagine actuaries using our research to decide on savings limits for participation in newly established pooled annuity funds.

%-----------------------
\printbibliography%biblatex uses different command to insert bibliography
%-----------------------

%----------------------------------------------
\appendix
\section*{Appendix}%no A in the section title Appendix
%----------------------------------------------
\section{Additional statements and proofs}
%----------------------------------------------

\counterwithin{defi}{section}
\setcounter{defi}{0}

\begin{lem}\label{lem:W_i/W_j=s_i/s_j}With the notation of Section \ref{section:operation}, at all integer times $\,t=0,1,2,\,$etc,
    \begin{enumerate}[label=(\alph*)]
        \item 
            $W_i(t)/W_j(t)=s_i/s_j\,$ for all $\,t<\min\{T_i,T_j\}$,
        \item         
            $\displaystyle M_i(t+1)=(W_i(t)-C_i(t))(1+R)\mathbbm{1}_{\{T_i>t+1\}}\frac{\sum_{k=1}^Ns_k\mathbbm{1}_{\{T_k\in(t,t+1]\}}}{\sum_{k=1}^Ns_k\mathbbm{1}_{\{T_k>t+1\}}}$.
    \end{enumerate}
\begin{proof}
    (a) Consider $\,t+1<\min\{T_i,T_j\}$. Define 
    \begin{equation*}
        \kappa(t+1)=\frac{\sum_{k=1}^N(W_k(t)-C_k(t))(1+R)\mathbbm{1}_{\{T_k\in(t,t+1]\}}}{\sum_{k=1}^N(W_k(t)-C_k(t))(1+R)\mathbbm{1}_{\{T_k>t+1\}}},
    \end{equation*}
    which is a finite non-negative random variable. Then,
    \begin{align*}
        \frac{W_i(t+1)}{W_j(t+1)}&\stackrel{(\ref{eq:W_i})}{=}\frac{(W_i(t)-C_i(t))(1+R)+M_i(t+1)}{(W_j(t)-C_j(t))(1+R)+M_j(t+1)}
        \\&\stackrel{(\ref{eq:D}),(\ref{eq:M_ioriginal})}{=}\frac{(W_i(t)-C_i(t))(1+R)(1+\,\kappa(t+1))}{(W_j(t)-C_j(t))(1+R)(1+\,\kappa(t+1))}
        \\&\stackrel{(\ref{eq:C_i})}{=}\frac{W_i(t)(1-1/\ddot{a}(x+t))(1+R)(1+\,\kappa(t+1))}{W_j(t)(1-1/\ddot{a}(x+t))(1+R)(1+\,\kappa(t+1))}=\frac{W_i(t)}{W_j(t)}.
    \end{align*}
    Using induction over $t$ together with $\,W_i(0)/W_j(0)=s_i/s_j\,$ yields the first statement.
    
    (b) Consider without loss of generality $\,T_i>t+1\,$ because the equation is otherwise trivial. Then, $\,W_i(t)>0\,$ and it follows that 
    \begin{align*}\label{eq:M_i}
        M_i(t+1)&\stackrel{(\ref{eq:D}),(\ref{eq:M_ioriginal})}{=}(W_i(t)-C_i(t))(1+R)\frac{\sum_{k=1}^N(W_k(t)-C_k(t))(1+R)\mathbbm{1}_{\{T_k\in(t,t+1]\}}}{\sum_{k=1}^N(W_k(t)-C_k(t))(1+R)\mathbbm{1}_{\{T_k>t+1\}}}
        \\&\stackrel{(\ref{eq:C_i})}{=}(W_i(t)-C_i(t))(1+R)\frac{\sum_{k=1}^NW_k(t)\mathbbm{1}_{\{T_k\in(t,t+1]\}}}{\sum_{k=1}^NW_k(t)\mathbbm{1}_{\{T_k>t+1\}}}
        \\&=(W_i(t)-C_i(t))(1+R)\frac{\sum_{k=1}^NW_k(t)/W_i(t)\mathbbm{1}_{\{T_k\in(t,t+1]\}}}{\sum_{k=1}^NW_k(t)/W_i(t)\mathbbm{1}_{\{T_k>t+1\}}}
        \\&\stackrel{(a)}{=}(W_i(t)-C_i(t))(1+R)\frac{\sum_{k=1}^Ns_k/s_i\mathbbm{1}_{\{T_k\in(t,t+1]\}}}{\sum_{k=1}^Ns_k/s_i\mathbbm{1}_{\{T_k>t+1\}}}
        \\&=(W_i(t)-C_i(t))(1+R)\frac{\sum_{k=1}^Ns_k\mathbbm{1}_{\{T_k\in(t,t+1]\}}}{\sum_{k=1}^Ns_k\mathbbm{1}_{\{T_k>t+1\}}},
    \end{align*} 
    which completes the second part of the proof.
\end{proof}
\end{lem}

\begin{pro}\label{pro:C_i(t)=C_i(0)}With the notation of Section \ref{section:operation}, at all integer times $\,t=0,1,2,\,$etc,
    \begin{enumerate}[label=(\alph*)]
        \item 
            $\displaystyle W_i(t)=\mathbbm{1}_{\{T_i>t\}}s_i\frac{\ddot{a}(x+t)}{\ddot{a}(x)}\frac{\actsymb[t]{p}{x}}{\actsymb[t]{\hat{p}}{x}}$,
        \item 
            $\displaystyle C_i(t)=\mathbbm{1}_{\{T_i>t\}}C_i(0)\frac{\actsymb[t]{p}{x}}{\actsymb[t]{\hat{p}}{x}}$.
    \end{enumerate}
\begin{proof}(a)
    \begin{align*}
        W_i(t+1)&\stackrel{(\ref{eq:W_i})}{=}\mathbbm{1}_{\{T_i>t+1\}}[(W_i(t)-C_i(t))(1+R)+M_i(t+1)]
        \\&\stackrel{\ref{lem:W_i/W_j=s_i/s_j}(b)}{=}\mathbbm{1}_{\{T_i>t+1\}}(W_i(t)-C_i(t))(1+R)(1+\frac{\sum_{k=1}^Ns_k\mathbbm{1}_{\{T_k\in(t,t+1]\}}}{\sum_{k=1}^Ns_k\mathbbm{1}_{\{T_k>t+1\}}})
        \\&\stackrel{(\ref{eq:tphatx})}{=}\mathbbm{1}_{\{T_i>t+1\}}(W_i(t)-C_i(t))(1+R)\frac{\actsymb[t]{\hat{p}}{x}}{\actsymb[t+1]{\hat{p}}{x}}
        \\&\stackrel{(\ref{eq:C_i})}{=}\mathbbm{1}_{\{T_i>t+1\}}W_i(t)(1-\frac{1}{\ddot{a}(x+t)})(1+R)\frac{\actsymb[t]{\hat{p}}{x}}{\actsymb[t+1]{\hat{p}}{x}}
        \\&\stackrel{(\ref{eq:ddota})}{=}\mathbbm{1}_{\{T_i>t+1\}}W_i(t)\frac{\ddot{a}(x+t+1)\actsymb[1]{p}{x+t}}{\ddot{a}(x+t)}\frac{\actsymb[t]{\hat{p}}{x}}{\actsymb[t+1]{\hat{p}}{x}}.
    \end{align*}
    Using the above formula inductively over time together with $\,\prod_{\delta=0}^{t-1}\actsymb[1]{p}{x+\delta}=\actsymb[t]{p}{x}$, $\,\actsymb[0]{\hat{p}}{x}=1$, and $\,W_i(0)=s_i$ yields
    \begin{align*}
        W_i(t)
        &=W_i(0)\prod_{\delta=0}^{t-1}\mathbbm{1}_{\{T_i>\delta+1\}}\frac{\ddot{a}(x+\delta+1)\actsymb[1]{p}{x+\delta}}{\ddot{a}(x+\delta)}\frac{\actsymb[\delta]{\hat{p}}{x}}{\actsymb[\delta+1]{\hat{p}}{x}}
        \\&=\mathbbm{1}_{\{T_i>t\}}s_i\frac{\ddot{a}(x+t)}{\ddot{a}(x)}\frac{\actsymb[t]{p}{x}}{\actsymb[t]{\hat{p}}{x}},
    \end{align*}
    which completes the proof of the first part of the statement.

    (b) is a consequence of (a) and (\ref{eq:C_i}).
\end{proof}
\end{pro}

\begin{pro}\label{ap-pro:Wealth-Donsker}
    Let $(s_i)_{i=1}^\infty$ be an i.i.d.\ sequence of strictly positive random variables with finite fourth moment. Let $(U_i)_{i=1}^\infty$ be an i.i.d.\ sequence of uniform random variables independent of $(s_i)_{i=1}^\infty$. Consider the process.
    \[X_N(u)=\frac{1}{\sqrt{\sum_{i=1}^Ns_i^2}}\sum_{i=1}^Ns_i(u-\mathbbm{1}_{\{U_i\leq u\}})\quad\mbox{for $\,N\in\mathbb{N},\,u\in[0,1]$}.\] 
    Then, the sequence $(X_N)_{N=1}^\infty$ converges conditional on $(s_i)_{i=1}^\infty$ in distribution in the Skorokhod space to a standard Brownian Bridge.
\begin{proof}
    Consider the regular conditional probability $\mathcal{P}$ of $(U_i)_{i=1}^\infty$ given $(s_i)_{i=1}^\infty$. Note that a countable number of equations are enough to describe that real-valued random variables are independent and uniformly distributed. Thus, because $(U_i)_{i=1}^\infty$ and $(s_i)_{i=1}^\infty$ are independent, $(U_i)_{i=1}^\infty$ is a sequence of i.i.d.\ uniform random variables for $\mathbb{P}$-almost-all outcomes of $(s_i)_{i=1}^\infty$ under $\mathcal{P}$. Hence, it is enough to show that $(X_N)_{N=1}^\infty$ converges in distribution to a standard Brownian Bridge for all fixed deterministic sequences $(s_i)_{i=1}^\infty$ that fulfil
    \begin{align}\label{eq:limsums/n}
        \quad\lim_{N\rightarrow\infty}\frac{\sum_{i=1}^Ns_i^2}{N}=\mu_2>0,
        \quad\lim_{N\rightarrow\infty}\frac{\sum_{i=1}^Ns_i^4}{N}=\mu_4.
    \end{align}

    We show that $(X_N)_{N=1}^\infty$ converges in distribution to a Brownian Bridge $X$ for any fixed $(s_i)_{i=1}^\infty$ with (\ref{eq:limsums/n}) by checking the following assumptions of Theorem 13.5 in \textcite[p.142]{Billingsley1999}:
    \begin{enumerate}[label=(\alph*)]
        \item The finite-dimensional distributions of $(X_N)_{N=1}^\infty$ converge to the ones of $X$,
        \item $X(1)-X(u)\xrightarrow[]{d}0\,$ for $\,u\rightarrow1$,
        \item There are constants $\,\gamma>0\,$ and $\,\alpha>1\,$ and a nondecreasing continuous function $f$ on $[0,1]$ such that for all large enough $\,N\in\mathbb{N}\,$ the following inequality holds for all $\,u\leq v\leq w\in[0,1]$, 
        \[\mathbb{E}\big[(X_N(w)-X_N(v))^\gamma(X_N(v)-X_N(u))^\gamma\big]\leq(f(w)-f(u))^\alpha.\]
    \end{enumerate}
    
    First, we establish that all linear combinations of marginals of $X_N$ fulfil Lyapunov's condition and hence converge to a normal distribution, see \textcite[p.362]{Billingsley1995}. Let $(u_j)_{j=1}^m$ and $(a_j)_{j=1}^m$ be finitely many points from $[0,1]$ and $\mathbb{R}$. Consider the following sequence of independent centred random variables
    \begin{align*}
        Y_i=s_i\sum_{j=1}^ma_j(u_j-\mathbbm{1}_{\{U_i\leq u_j\}})\quad\mbox{for $\,i\in\mathbb{N}$}.
    \end{align*}
    Because $(U_i)_{i=1}^\infty$ is a sequence of identical distributed uniform random variables, there are finite constants $\kappa_2$ and $\kappa_4$ depending on $(u_j)_{j=1}^m$ and $(a_j)_{j=1}^m$ but independent from $i$ such that
    \begin{align*}
        \mathbb{E}[Y_i^2]=s_i^2\,\mathbb{E}\big[\big(\sum_{j=1}^ma_j(u_j-\mathbbm{1}_{\{U_1\leq u_j\}})\big)^2\big]=s_i^2\,\kappa_2,
        \\\mathbb{E}[Y_i^4]=s_i^4\,\mathbb{E}\big[\big(\sum_{j=1}^ma_j(u_j-\mathbbm{1}_{\{U_1\leq u_j\}})\big)^4\big]=s_i^4\,\kappa_4.
    \end{align*}
    Hence, in view of (\ref{eq:limsums/n}),
    \begin{align*}
        \frac{\sum_{i=1}^N\mathbb{E}[Y_i^4]}{(\sum_{i=1}^N\mathbb{E}[Y_i^2])^2}
        =\frac{\sum_{i=1}^Ns_i^4\,\kappa_4}{(\sum_{i=1}^Ns_i^2\,\kappa_2)^2}
        =\frac{\kappa_4}{N\kappa_2^2}\frac{\sum_{i=1}^Ns_i^4}{N}\big(\frac{N}{\sum_{i=1}^Ns_i^2}\big)^2\xrightarrow[]{N\uparrow\infty}0\times\mu_4\times\frac{1}{\mu_2^2}=0.
    \end{align*}
    Thus, $(Y_i)_{i=1}^\infty$ is a sequence of independent centered random variables that fulfils Lyapunov's condition and therefore $\,(\sum_{i=1}^NY_i/(\kappa_2\sum_{i=1}^Ns_i^2)^{1/2})_{N=1}^\infty$ converges to a normal distribution. Hence, the linear combination $\,\sum_{j=1}^ma_jX_N(u_j)=\sum_{i=1}^NY_i/(\sum_{i=1}^Ns_i^2)^{1/2}\,$ converges for $\,N\uparrow\infty\,$ to a normal distribution. Because $(u_j)_{j=1}^m$ and $(a_j)_{j=1}^m$ are arbitrary, all linear combinations of marginals of $X_N$ converge to a normal distribution.
    
    Next, we confirm that the means and covariances of the marginals converge to the ones of a Brownian bridge. Since $\,u-\mathbb{E}[\mathbbm{1}_{U_i\leq u}]=0\,$ and $(U_i)_{i=1}^\infty$ i.i.d., we have for $\,u\leq v\,$ from $[0,1]$ that
    \begin{align*}
        \mathbb{E}[X_N(u)]&=\frac{1}{\sqrt{\sum_{i=1}^Ns_i^2}}\sum_{i=1}^Ns_i(u-\mathbb{E}[\mathbbm{1}_{U_i\leq u}])=0,
        \\\mathrm{Cov}(X_N(u),X_N(v))&=\mathbb{E}[(u-\mathbbm{1}_{U_1\leq u})(v-\mathbbm{1}_{U_1\leq v})]
        \\&=uv-\mathbb{E}[\mathbbm{1}_{U_1\leq v}]u-\mathbb{E}[\mathbbm{1}_{U_1\leq u}]v+\mathbb{E}[\mathbbm{1}_{U_1\leq u}]
        \\&=u(1-v).
    \end{align*}
    Hence, the finite-dimensional distributions of $(X_N)_{N=0}^\infty$ converge to the ones of a Brownian Bridge, which shows that assumption (a) holds.
    
    Assumption (b) holds automatically because the paths of a standard Brownian Bridge are continuous throughout its domain.
    
    Fix $\,u\leq v\leq w\,$ from $[0,1]$ and consider
    \begin{align*}
        Z_i&=(w-v)-\mathbbm{1}_{v<U_i\leq w},
        \\\tilde{Z}_i&=(v-u)-\mathbbm{1}_{u<U_i\leq v}.
    \end{align*}
    Note that $(w-v)-\mathbb{E}[\mathbbm{1}_{v<U_i\leq w}]=(v-u)-\mathbb{E}[\mathbbm{1}_{u<U_i\leq v}]=0\,$ and $(U_i)_{i=1}^\infty$ is a sequence of i.i.d.\ uniformly distributed random variables, hence,
    \begin{gather}
        \nonumber\mathbb{E}\big[(X_N(w)-X_N(v))^2(X_N(v)-X_N(u))^2\big]
        =\frac{1}{(\sum_{i=1}^Ns_i^2)^2}\sum_{i,j,k,\ell=1}^Ns_is_js_ks_\ell\,\mathbb{E}[Z_iZ_j\tilde{Z}_k\tilde{Z}_\ell]
        \\\label{eq:E[(X-X)(X-X)]}=\frac{1}{(\sum_{i=1}^Ns_i^2)^2}\Big(\mathbb{E}[Z_1^2\tilde{Z}_1^2]\sum_{i=1}^Ns_i^4+\mathbb{E}[Z_1^2]\mathbb{E}[\tilde{Z}_1^2]\sum_{i\neq j}s_i^2s_j^2+\mathbb{E}^2[Z_1\tilde{Z}_1]\sum_{i\neq k}2s_i^2s_k^2\Big).
    \end{gather}
    The following estimate is taken from \textcite[p.150]{Billingsley1999},
    \begin{align}\label{eq:maxEEE}
        \max\big\{\mathbb{E}[Z_1^2\tilde{Z}_1^2],\mathbb{E}[Z_1^2]\mathbb{E}[\tilde{Z}_1^2],\mathbb{E}^2[Z_1\tilde{Z}_1]\big\}\leq3(w-u)^2.
    \end{align}
    Moreover, using (\ref{eq:limsums/n}) again, we find that
    \begin{align}\label{eq:sumsi^4/()^2}
        \frac{\sum_{i=1}^Ns_i^4}{(\sum_{i=1}^Ns_i^2)^2}
        &=\frac{1}{N}\frac{\sum_{i=1}^Ns_i^4}{N}\big(\frac{N}{\sum_{i=1}^Ns_i^2}\big)^2\xrightarrow[]{N\uparrow\infty}0\times\mu_4\times\frac{1}{\mu_2^2}=0,
        \\\label{eq:sumsi^2sj^2/()^2}
        \frac{\sum_{i\neq j}s_i^2s_j^2}{(\sum_{i=1}^Ns_i^2)^2}
        &\leq\frac{\sum_{i,j=1}^Ns_i^2s_j^2}{(\sum_{i=1}^Ns_i^2)^2}
        =\frac{(\sum_{i=1}^Ns_i^2)(\sum_{j=1}^Ns_j^2)}{(\sum_{i=1}^Ns_i^2)^2}=1.
    \end{align}
    Combining (\ref{eq:E[(X-X)(X-X)]}), (\ref{eq:maxEEE}), (\ref{eq:sumsi^4/()^2}), (\ref{eq:sumsi^2sj^2/()^2}) and taking $N$ large enough so that the estimate in (\ref{eq:sumsi^4/()^2}) is smaller than $1$, which can be chosen for all values of $u,v$ and $w$, yields
    \[\mathbb{E}\big[(X_N(w)-X_N(v))^2(X_N(v)-X_N(u))^2\big]\leq12(w-u)^2,\]
    which implies (c). 
    
    Overall, $(X_N)_{N=1}^\infty$ converges in distribution to a standard Brownian Bridge.
\end{proof}
\end{pro}

\begin{pro}\label{pro:>N4mM/(m+M)Part2}
    Consider the space $\mathcal{S}$ of savings $\,s=(s_i)_{i=1}^N$ with fixed number of members $N$ and fixed bounds $\,m\leq s_i\leq M\,$ for all $\,i\,$ where $\,M>m>0$. Then
    \begin{equation*}
        \inf_{s\in\mathcal{S}}\nu(s)\leq N\frac{4mM}{(m+M)^2}(1+\frac{M^3}{4m^3N^2}).
    \end{equation*}
\begin{proof}
    Using the same notation as in the proof of Proposition \ref{pro:>N4mM/(m+M)}, the uniqueness of $x^*$ implies that $f$ is decreasing on $[0,x^*]$ and increasing on $[x^*,1]$. Thus, the optimal integer $n^*$ that achieves (\ref{eq:inf=Bin}) fulfils
    \begin{equation}\label{eq:n*/N}
        \frac{n^*}{N}=\frac{m}{M+m}+\delta\quad\mbox{with $\,|\delta|\leq\frac{1}{N}$}.
    \end{equation}
    Plugging (\ref{eq:n*/N}) into (\ref{eq:inf=Bin}) and using the definition $\,\varepsilon=(M^2-m^2)/(Mm)\,$ yields
    \begin{equation}\label{eq:inf=(delta)}
        \inf_{s\in\mathcal{S}}\nu(s)=N\frac{4mM}{(m+M)^2}(1+\frac{\varepsilon^2\delta^2}{4(1+\varepsilon\delta)}).
    \end{equation}
    Next, we estimate $\,1+\varepsilon\delta$. Note that $\,\delta\geq0\,$ implies $\,\varepsilon\delta\geq0\,$ and that $\,\delta<0\,$ implies $\,|\delta|\leq m/(M+m)\,$ because $0$ is attainable for $n/N$ in (\ref{eq:inf=Bin}). Hence, 
    \begin{align*}
        \delta\geq0:&\quad1+\varepsilon\delta\geq1,
        \\\delta<0:&\quad1+\varepsilon\delta=  1-\varepsilon|\delta|=1-\frac{(M-m)(M+m)}{mM}|\delta|\geq1-\frac{M-m}{M}=\frac{m}{M}.
    \end{align*}
    Overall, we can see that $\,1+\varepsilon\delta\geq m/M$. Furthermore, $\,\varepsilon=(M^2-m^2)/(Mm)\leq M^2/(Mm)=M/m$. Those two estimates applied to (\ref{eq:inf=(delta)}) yields
    \begin{equation*}
        \inf_{s\in\mathcal{S}}\nu(s)=N\frac{4mM}{(m+M)^2}(1+\frac{\varepsilon^2\delta^2}{4(1+\varepsilon\delta)})\leq N\frac{4mM}{(m+M)^2}(1+\frac{M^3}{4m^3N^2}).
    \end{equation*}
\end{proof}
\end{pro}

%----------------------------------------------
\section{Program design}
%----------------------------------------------
\renewcommand\qedsymbol{}% to remove qed-symbol

\begin{simu}\label{simu:2groups}
    We use $\,R\,$ Monte Carlo simulations to find the maximal $\,u=F(t)\in[0,1]\,$ such that $t$ fulfills (\ref{eq:P>beta}) where $F$ is the distribution function of an underlying mortality distribution. (\ref{eq:P>beta}) depends on the parameters $\,\varepsilon_1\in(0,1),\,\varepsilon_2>0$ and $\,\beta\in[0,1]$. We specifically consider a group of $N$ members split into two subgroups. The first subgroup has $N_1$ members and each member has the same initial savings, i.e.\ $m=1\,$ for $\,1\leq i\leq N_1$. The second subgroup has $\,N-N_1$ members and each member has the same initial savings, possibly different from the first group, i.e.\ $\,M=a\,$ for $\,i>N_1$.
    
\begin{proof}[Remark]
    The value $u$ is independent of $F$. To see this, consider $\,(U_i)_{i=1}^N=(F(T_i))_{i=1}^N$, which is a sequence of i.i.d.\ uniform random variables. Then, let $\,v\in[0,1]$, and observe that
    \begin{gather*}
        \actsymb[F^{-1}(v)]{p}{x}=1-v,
        \\\actsymb[F^{-1}(v)]{\hat{p}}{x}=1-\frac{1}{\sum_{j=1}^Ns_j}\sum_{i=1}^Ns_i\mathbbm{1}_{\{U_i\leq v\}}=\colon 1-\hat{F}(F^{-1}(v)).
    \end{gather*}
\end{proof}

\begin{proof}[Steps/Description]    
    \begin{enumerate}
        \item
            Generate the order statistic $(U_{(k)})_{k=1}^{N}$ of $N$ independent and uniformly distributed random variables, use exponential random variables for a fast simulation, see \textcite[ff.207]{Devroye1986}. 
            \\\hspace*{10pt}The order statistic represents the increasing and $F$-transformed lifetimes of the group. 
        \item   
            Choose $N_1$ numbers at random from $1$ to $N$ without replacement, e.g.\ the first $N_1$ numbers of a random permutation of $N$ elements. 
            \\\hspace*{10pt}The numbers indicate which lifetimes belong to members of the first subgroup.
        \item
            Find the first $k$ between $0$ and $N$ such that
            \begin{align}
                 \label{eq:<1-eps}\frac{1-U_{(k+1)}}{1-\hat{F}(F^{-1}(U_{(k)}))}&< 1-\varepsilon_1
                 \quad\mbox{or}
                 \\\label{eq:1+eps<}1+\varepsilon_2&<\frac{1-U_{(k)}}{1-\hat{F}(F^{-1}(U_{(k)}))}
            \end{align}
            with $\,U_{(0)}=0\,$ and $\,U_{(N+1)}=1$. 
            \\\hspace*{10pt}The integer $k$ indicates the last member who receives a lifelong stable income under Definition \ref{defi:P>beta}(ii). To see this, observe that $\,v\mapsto(1-v)/(1-\hat{F}(F^{-1}(v)))\,$ is decreasing between consecutive $U_{(k)}$. Thus, if $\,(1-v)/(1-\hat{F}(F^{-1}(v)))<1-\varepsilon_1\,$ holds within $\,[U_{(k)},U_{(k+1)})$ for the first time, then $\,(1-v)/(1-\hat{F}(F^{-1}(v)))<1-\varepsilon_1\,$ holds also at the right limit of the intervals $\,[U_{(k)},U_{(k+1)})$ for the first time, and vice versa. Similarly, $\,(1-v)/(1-\hat{F}(F^{-1}(v)))>1+\varepsilon_2\,$ holds for the first time at left end points of the intervals $\,[U_{(k)},U_{(k+1)})$, and vice versa.
        \item 
            Compute 
            \begin{equation*}
                \tau=
                \begin{cases}
                     \hfill U_{(k)}&\quad\mbox{if $\,1+\varepsilon_2<\frac{1-U_{(k)}}{1-\hat{F}(F^{-1}(U_{(k)}))}$}
                     \\
                     1-(1-\varepsilon_1)(1-\hat{F}(F^{-1}(U_{(k)})))&\quad\mbox{otherwise}
                \end{cases}
            \end{equation*}
            \\\hspace*{10pt}The random variable $\tau$ represents the $F$-transformed time when the pooled annuity fund stops to pay a stable income under Definition \ref{defi:P>beta}(ii). To see this, observe that $k$ is the last member who receives a lifelong stable income. Hence, the income is unstable for the first time within $\,[U_{(k)},U_{(k+1)})$. If (\ref{eq:1+eps<}) holds, then trivially the income is unstable at $U_{(k)}$, the earliest possible point in $\,[U_{(k)},U_{(k+1)})$. Hence, $U_{(k)}$ is the first time the income becomes unstable. On the other hand, if (\ref{eq:1+eps<}) fails, then (\ref{eq:<1-eps}) holds. As $k$ is minimal and only upward jumps occur, it holds that
            \begin{equation}\label{eq:1-eps<left}
                1-\varepsilon_1<\frac{1-U_{(k)}}{1-\hat{F}(F^{-1}(U_{(k-1)}))}<\frac{1-U_{(k)}}{1-\hat{F}(F^{-1}(U_{(k)}))}.
            \end{equation}
            In particular, (\ref{eq:<1-eps}) and (\ref{eq:1-eps<left}) guarantee that there is a solution to $\,(1-v)/(1-\hat{F}(F^{-1}(v)))=1-\varepsilon_1$ in $\,[U_{(k)},U_{(k+1)})$, which we find easily as $\,\hat{F}(F^{-1}(v)))\,$ is constant in $\,[U_{(k)},U_{(k+1)})$.
        \item 
            Create $R$ samples of $\tau$ (repeat steps 1.\ to 4.) and find the $(1-\beta)$-quantile of the samples.
    \end{enumerate}
\end{proof}
\end{simu}

\end{document}